\documentclass{egpubl}

 \electronicVersion 

\ifpdf \usepackage[pdftex]{graphicx} \pdfcompresslevel=9
\else \usepackage[dvips]{graphicx} \fi

\PrintedOrElectronic

\usepackage{t1enc,dfadobe}

\usepackage{egweblnk}

\title%
      {The State of the Art in Cartograms
}

\author[S. Nusrat  and S. Kobourov]
{
Sabrina~Nusrat and 
Stephen~Kobourov\\
Department of Computer Science, University of Arizona
}

\begin{document}


\maketitle

\begin{abstract}
   Cartograms combine statistical and geographical information in thematic maps, where areas of geographical regions (e.g., countries, states) are scaled in proportion to some statistic (e.g., population, income).  Cartograms make it possible to gain insight into patterns and trends in the world around us and have been very popular visualizations for geo-referenced data for over a century. 
   This work surveys cartogram research in visualization, cartography and geometry, covering a broad spectrum of different cartogram types: from the traditional rectangular and table cartograms, to Dorling and diffusion cartograms. A particular focus is the study of the major cartogram dimensions: statistical accuracy, geographical accuracy, and topological accuracy. We review the history of cartograms, describe the algorithms for generating them, and consider task taxonomies. We also review quantitative and qualitative evaluations, and we use these to arrive at design guidelines and research challenges.


\end{abstract}

\section{Introduction}
Cartograms combine statistical and geographical information in thematic maps, where areas of geographical regions (e.g., countries, states) are scaled in proportion to some statistic (e.g., population, income). 
This kind of visualization has been used for many years, with the first reference to the term ``cartogram'' dating back to at least 1870. 
Since then, cartograms have been studied by geographers, cartographers, economists, social scientists, geometers, and information visualization researchers. Many different types of cartograms have been proposed and implemented, optimizing different aspects, such as statistical accuracy (cartographic error), geographical accuracy (preserving the outlines of geographic shapes), and topological accuracy (maintaining correct adjacencies between countries).

Since cartograms combine statistical and geographical information, they can provide insight into patterns, trends and outliers in the world around us. Researchers in cartography, computational geometry and information visualization have designed, implemented and evaluated many different algorithms for generating cartograms.
Likely due to their aesthetic appeal, along with the possibility to combine geographical, political and socioeconomic data, cartograms are also widely used in newspapers, magazines, textbooks, blogs, and presentations. 

Even though several excellent cartogram surveys exist~\cite{guseyn1994numerical,kocmoud1997constructing,Tobler04}, more than a decade has passed since the last one. In the meantime, more than 70 new papers about cartograms have appeared in journals and conference proceedings on information visualization, cartography and computational geometry. Notably, several new cartogram models have been proposed since the most recent survey. 
Further, there has been a great deal of work on evaluating the broad spectrum of cartogram algorithms, both by performance measures and by subjective preference and user-studies. This warrants an attempt to reconsider all methods, to classify cartograms by design dimensions, and to analyze current trends and future directions.

  With this in mind, we survey cartogram research in visualization, cartography and geometry, covering a broad spectrum of different cartogram types: from rectangular and table cartograms, to Dorling and diffusion cartograms. A particular focus is the study of the major cartogram dimensions: statistical accuracy, geographical accuracy, and topological accuracy. 
We survey the historical use of cartograms, the cartogram literature, and describe the main techniques for generating cartograms. We also and review cartogram evaluation studies, task taxonomies, and make recommendations for the use of cartograms in different settings.

\section{Scope and Methodology}



We review the history and evolution of cartograms, from the early
hand-drawn examples to modern cartogram-generation algorithms,
taxonomies, evaluations, and applications on cartograms. We
focus mostly on {\em value-by-area} types of maps and thus many other
thematic maps, such as choropleth maps, graduated circle maps, and travel-distance maps, are beyond the scope of this survey. 

We created a database of research papers about cartograms using the SurVis system~\cite{beck2016visual}. 
Our literature dataset includes a wide range of publications: from aspects of cartogram computation to evaluations, perception and cognitive aspects of map reading. We populated the database starting with relevant citations from our research papers on cartogram-generation methods~\cite{ourSoCG,KKN13}, cartogram taxonomies~\cite{Task_C}, and evaluations~\cite{AKV15}, as well as manually inspecting the following journals and conference proceedings:


\begin{itemize}

\item Journals

-- \textit{IEEE Transactions on Visualization and Computer Graphics}

-- \textit{Computer Graphics Forum}

-- \textit{Information Visualization}

-- \textit{The American Cartographer}

-- \textit{The Cartographic Journal}


-- \textit{Discrete \& Computational Geometry}

\item Conferences

-- \textit{IEEE Symposium on Information Visualization (InfoVis)}

-- \textit{Eurographics VGTC Symposium on Visualization (EuroVis)}

-- \textit{IEEE Pacific Visualization Symposium (PacificVis)}

-- \textit{ACM Symposium on Computational Geometry (SoCG)}






\end{itemize}

\section{Origin and History of Cartograms}
\label{sec:early}

According to Tobler~\cite{Tobler04}, the first reference to the term ``cartogram'' dates back to 1870, when \'Emile Levasseur's cartograms were used in an economic geography textbook. 
According to Fabrikant~\cite{Fabrikant_commentary}, cartograms were used to depict German election results as early as in 1903. 
In 1934 Raisz~\cite{Raisz34}, gave the first formal definition of rectangular cartograms. 
In this section we attempt to put together a brief history of cartograms, starting with pre-twentieth century ideas.

\subsection{19th Century Cartograms}

\begin{figure}[htbp]
\centering
\includegraphics[width=.49\textwidth]{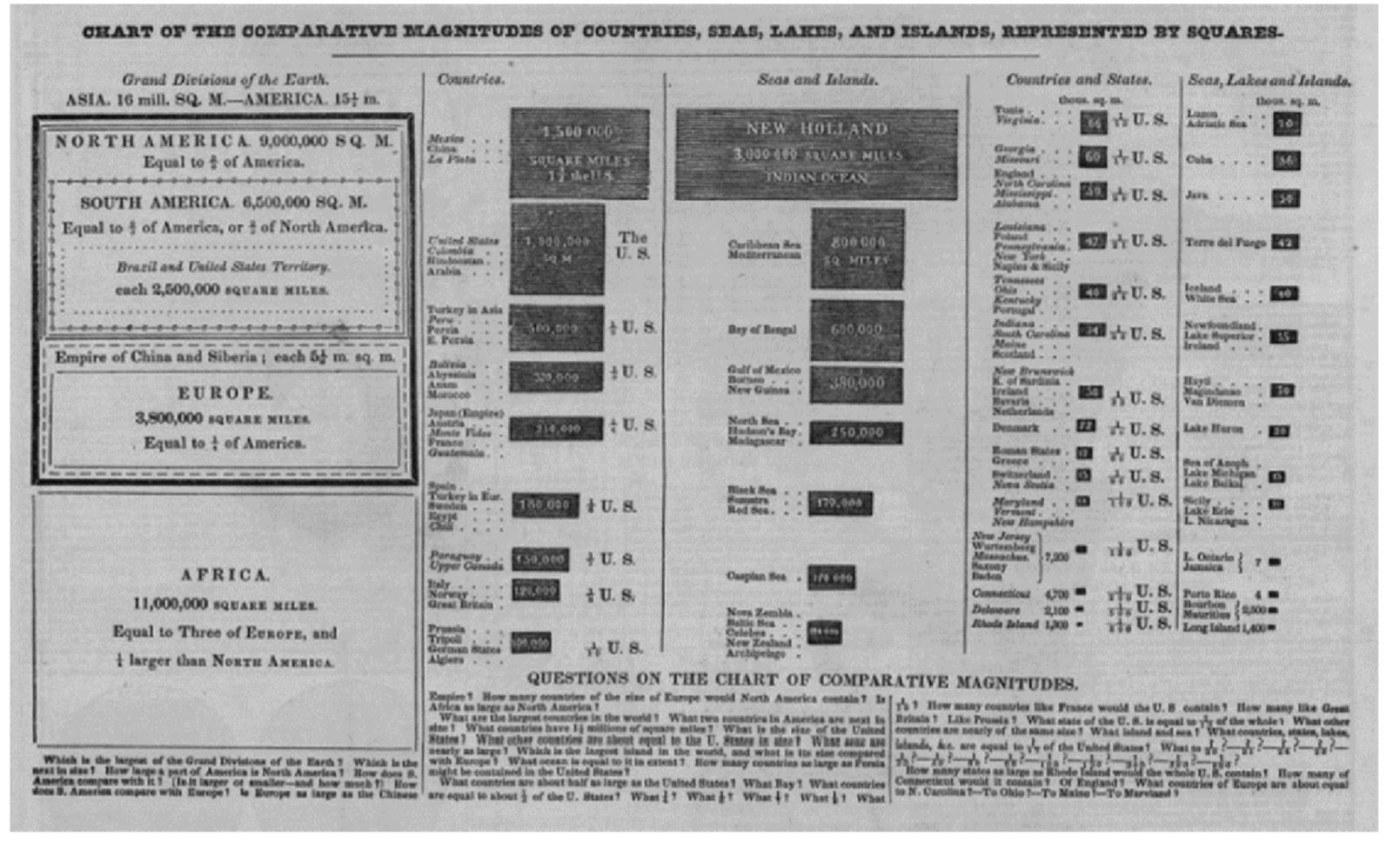}\\

(a) ``Chart of the comparative magnitudes of countries'' by Woodbridge in 1837.\\\vspace{.25cm}
\includegraphics[width=.49\textwidth]{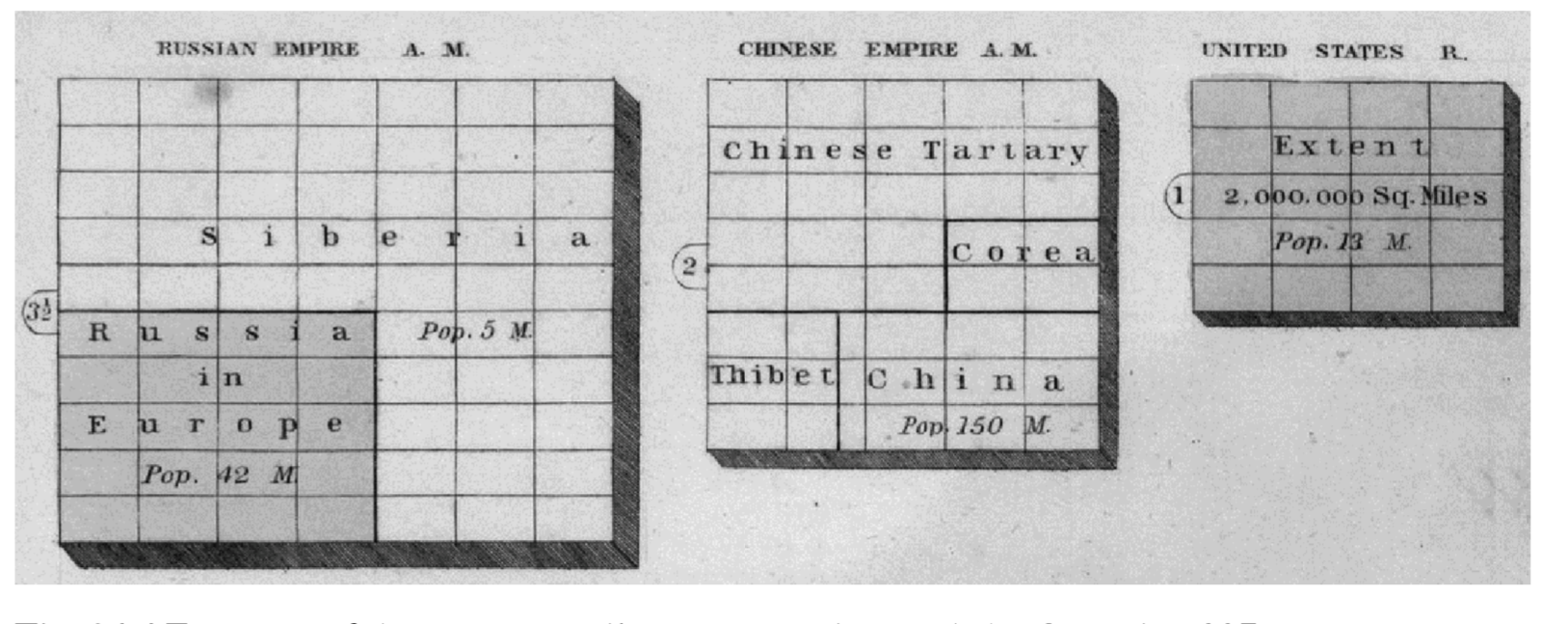}\\
(b) Figure from the ``New and Improved School Atlas'' by Olney in 1837.\\\vspace{.25cm}
\includegraphics[width=.49\textwidth]{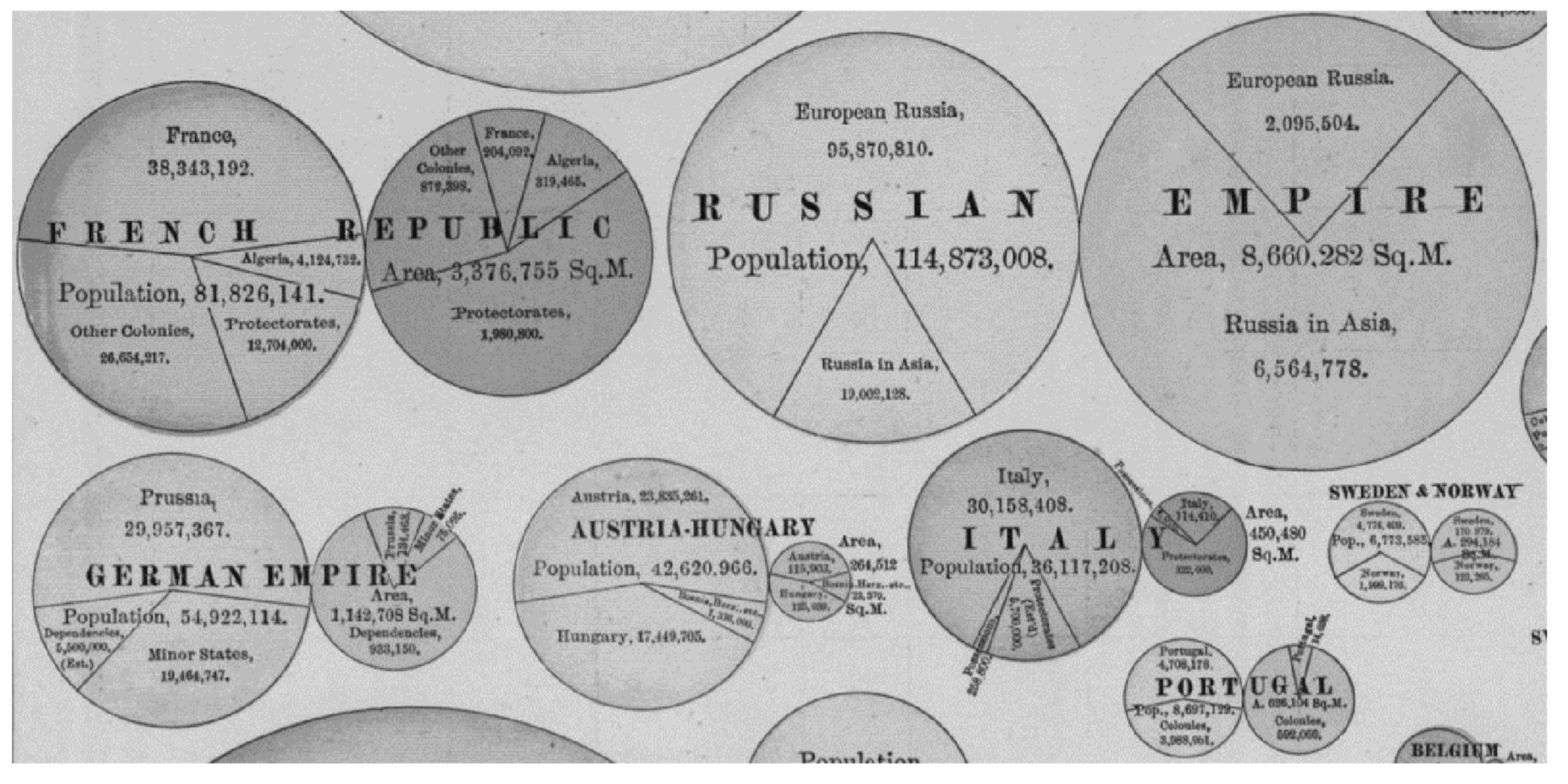}\\
(c) Figure from the Rand McNally World Atlas of 1897.\\\vspace{.25cm}
\includegraphics[width=.49\textwidth]{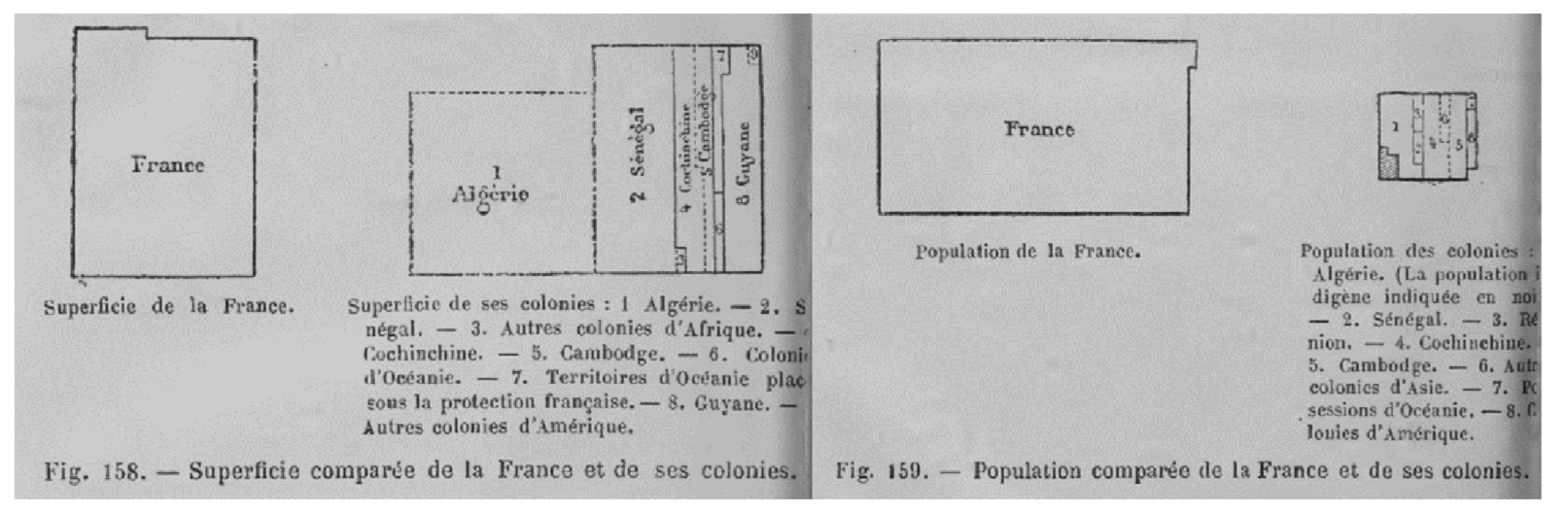}\\
(d) Illustration of geographical size and population of France and its colonies in a 1875 Geography textbook by \'Emile Levasseur.
 \caption{Cartogram-like representations from ``The use of cartograms in school cartography''~\cite{school_cartography}.
\label{fig:cartograms}}
\end{figure}

Cartogram-like representations can be found in 19th century atlases in the US. One of the earliest examples is published by William C. Woodbridge in 1837 in his ``Modern atlas, on a new plan, to accompany the system of Universal Geography'' where he used ``comparative charts'' of North America, Europe, Africa, and South America; see Fig.~\ref{fig:cartograms}(a). These charts were improved and reproduced in the 1843 ``Modern Atlas''  and in the 1845 ``School Atlas, to accompany Modern School Geography''. In 1837 Jesse Olney included a cartogram-like representation in his ``New and Improved School Atlas''  to show the size and population of the principal empires and kingdoms;
see Fig.~\ref{fig:cartograms}(b). The Rand McNally World Atlas of 1897 also published some cartogram-like representations. One such representation had two circles for each empire, one symbolizing the area of the empire, other symbolizing the population~\cite{school_cartography}; see Fig.~\ref{fig:cartograms}(c). 

Cartogram-like representations became popular in France in the 19th century, and were  published in newspapers, journals, and atlases~\cite{school_cartography}. The French economist, geographer and educator, Pierre \'Emile Levasseur, is considered a pioneer of the use of cartogram-like representations in school textbooks: Fig.~\ref{fig:cartograms}(d) shows an example taken from page 778 of his Geography textbook, ``La France, avec ses Colonies...'' published in Paris in 1875. 

\subsection{20th Century Cartograms}

\begin{figure}[t]
\begin{center}
\includegraphics[width=0.48\textwidth]{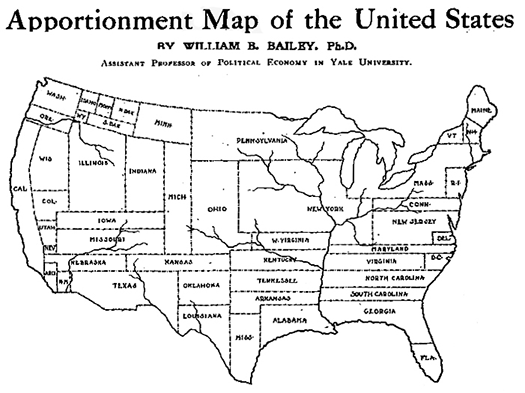}\\
\caption{1911 Apportionment Map of the United States by William B. Bailey, published in {\em The Independent}~\cite{john_carto}.}
\label{fig:bailey}
\end{center}
\end{figure}

Early in the 20th century, the term ``cartogram'' was inconsistently used to refer to various kinds of charts and maps, such as bar charts and graduated circle maps~\cite{Bailey_carto}; for example, Funkhouser~\cite{Funk} used the term ``cartogram''  to describe a choropleth map.
Figure~\ref{fig:bailey} shows William B. Bailey's 1911 
 ``Apportionment Map of the United States,'' which scales the size of the states according to their population~\cite{krygier2011making}.
 Bailey also provided an informal description of a population cartogram as follows~\cite{Bailey_carto}:

\hspace{0.01\textwidth}\parbox{0.45\textwidth}{\textit{``The map shown on this page is drawn on the principle that the population is evenly distributed throughout the whole United States, and that the area of the States varies directly with their population. With the map constructed on this principle some curious changes become apparent. On the ordinary map the four States, Montana, Wyoming, Colorado, and New Mexico, together with the seven States which lie to the west of them, comprise more than one-third of the territory of the United States, and the area of each one of them is considerably larger than that of New York State; yet the population of New York State alone is nearly one-fourth larger than the combined population of these eleven Western States.''}}

In 1934 Raisz~\cite{Raisz34} gave a more formal definition of value-by-area cartogram, specifying that:

\hspace{0.01\textwidth}\parbox{0.45\textwidth}{\textit{``the statistical cartogram is not a map. Although it has roughly the proportions of the country and retains as far as possible the relative locations of the various regions, the cartogram is purely a geometrical design to visualize certain statistical facts and to work out certain problems of distribution.''}}

\begin{figure}[htbp]
\begin{center}
\includegraphics[width=.42\textwidth]{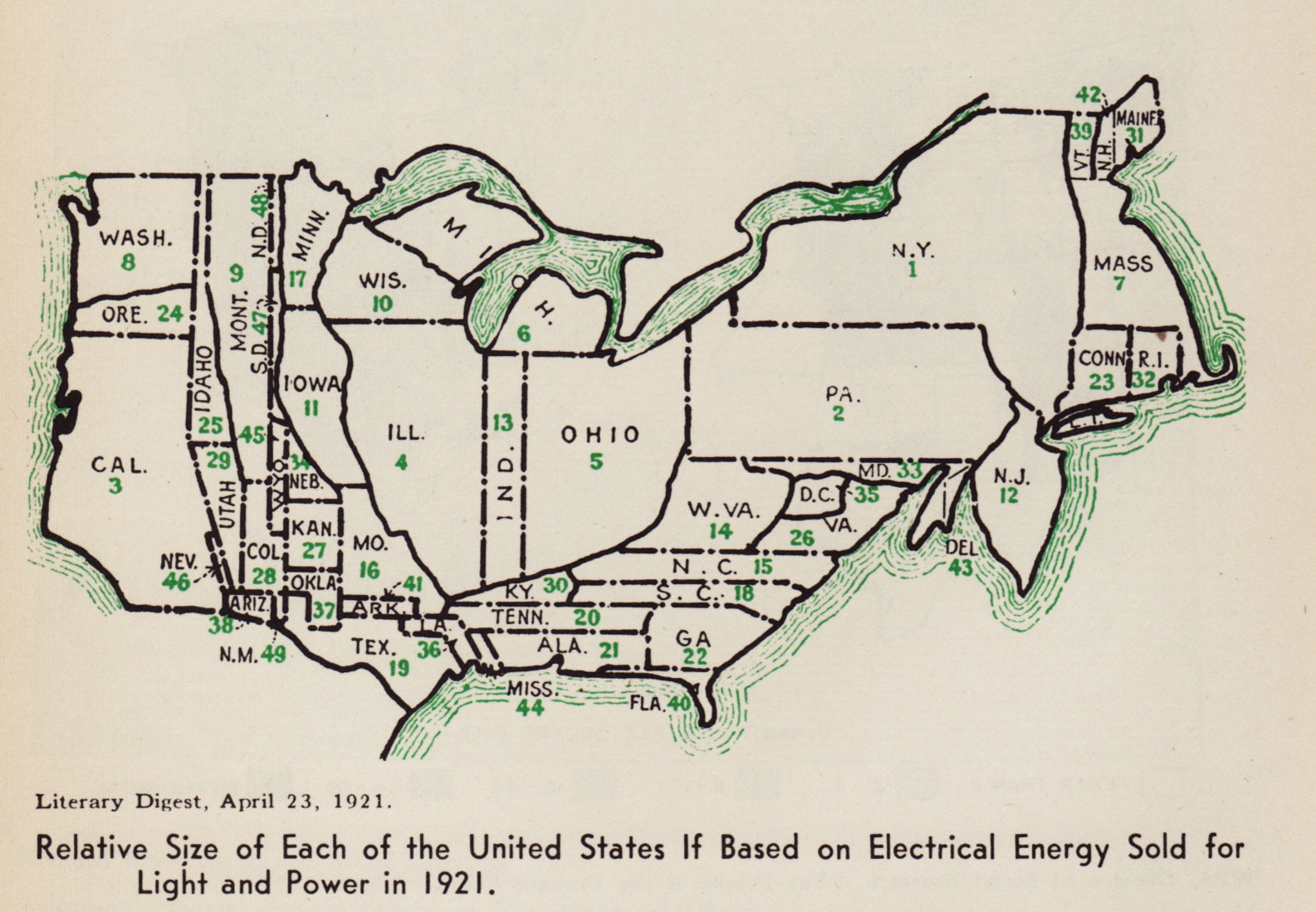}\\
(a) A 1921 cartogram of the USA based on electrical energy sold for light and power in the Literary Digest.\\\vspace{.2cm}
\includegraphics[width=0.45\textwidth]{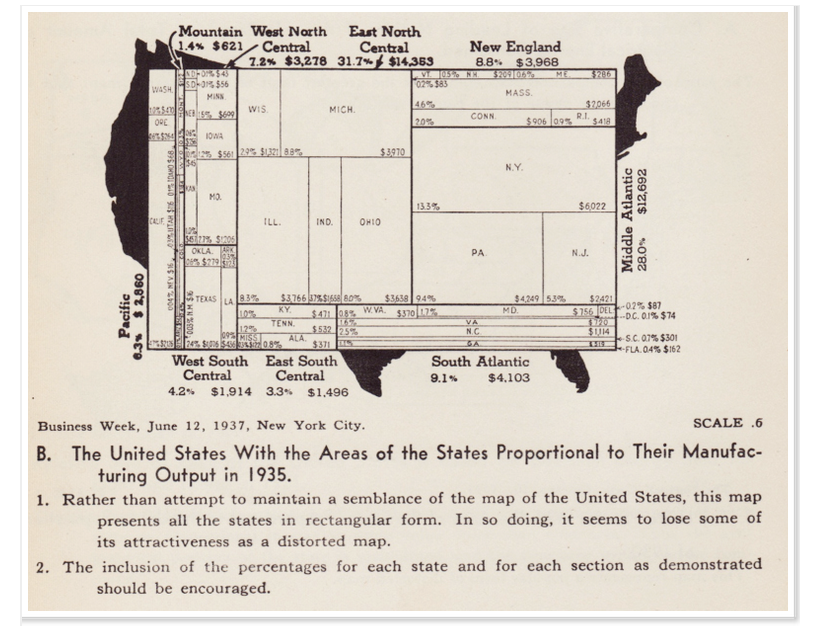}\\
(b) A 1937 cartogram of manufacturing output in Business Week.\\\vspace{.2cm}
\includegraphics[width=0.45\textwidth]{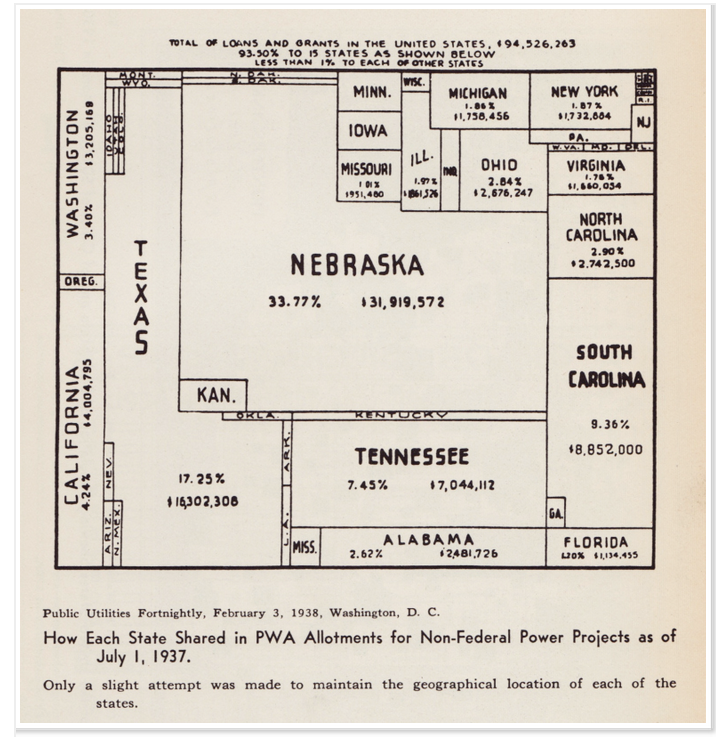}\\
(c) A 1938 cartogram of the Public Work Administration power projects in Public Utilities Fortnightly.\\

\caption{Early cartograms in US magazines from~\cite{john_carto}.}
\vspace{-0.2cm}
\label{fig:businessweek}
\end{center}
\end{figure}

\begin{figure}[htbp]
\centering
\vspace{.2cm}
\includegraphics[width=.49\textwidth]{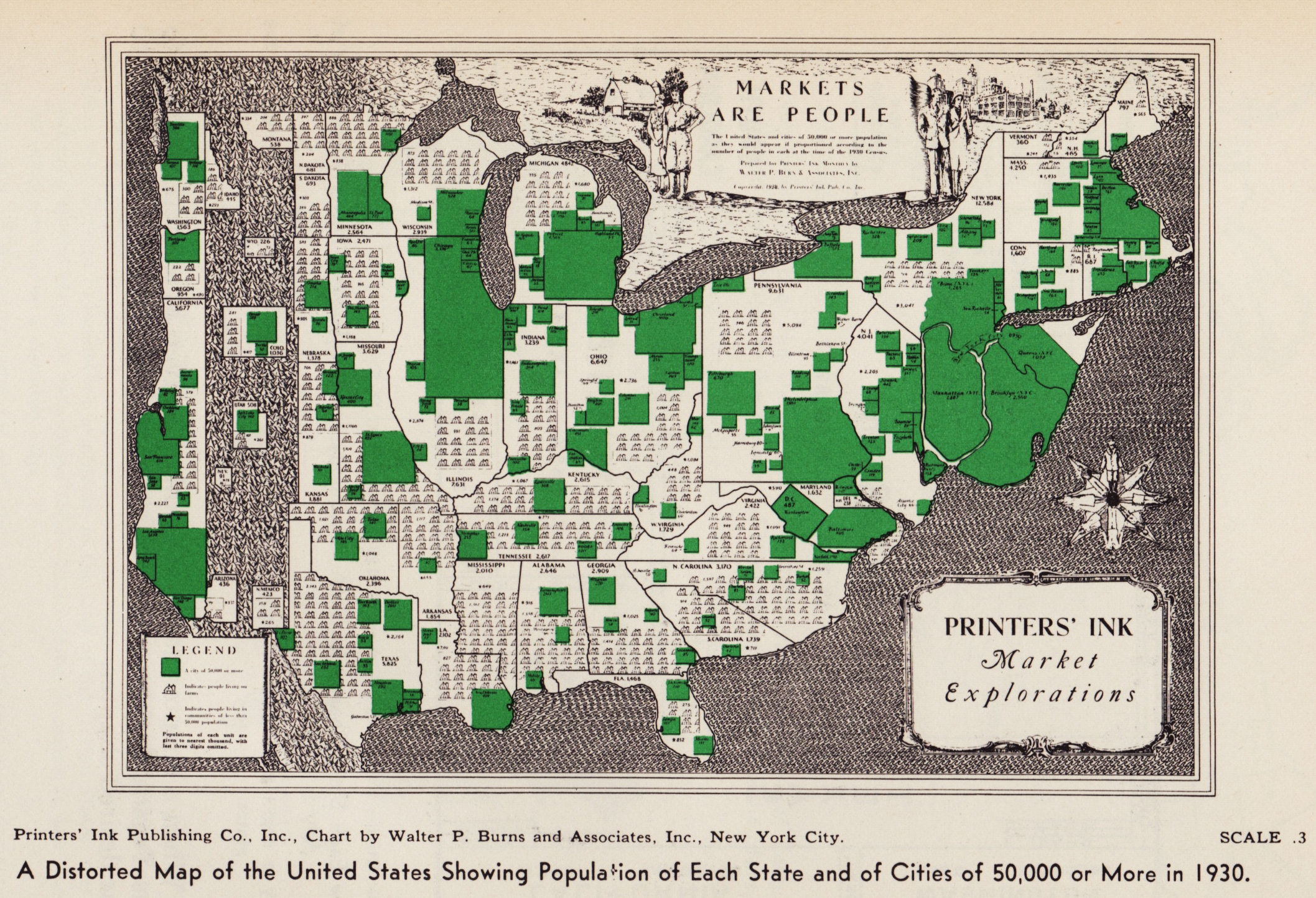}\\
(a) ``A Distorted Map of the United States Showing Population of Each State and of Cities of 50,000 or More in 1930.'' \\\vspace{.25cm}
\includegraphics[width=.49\textwidth]{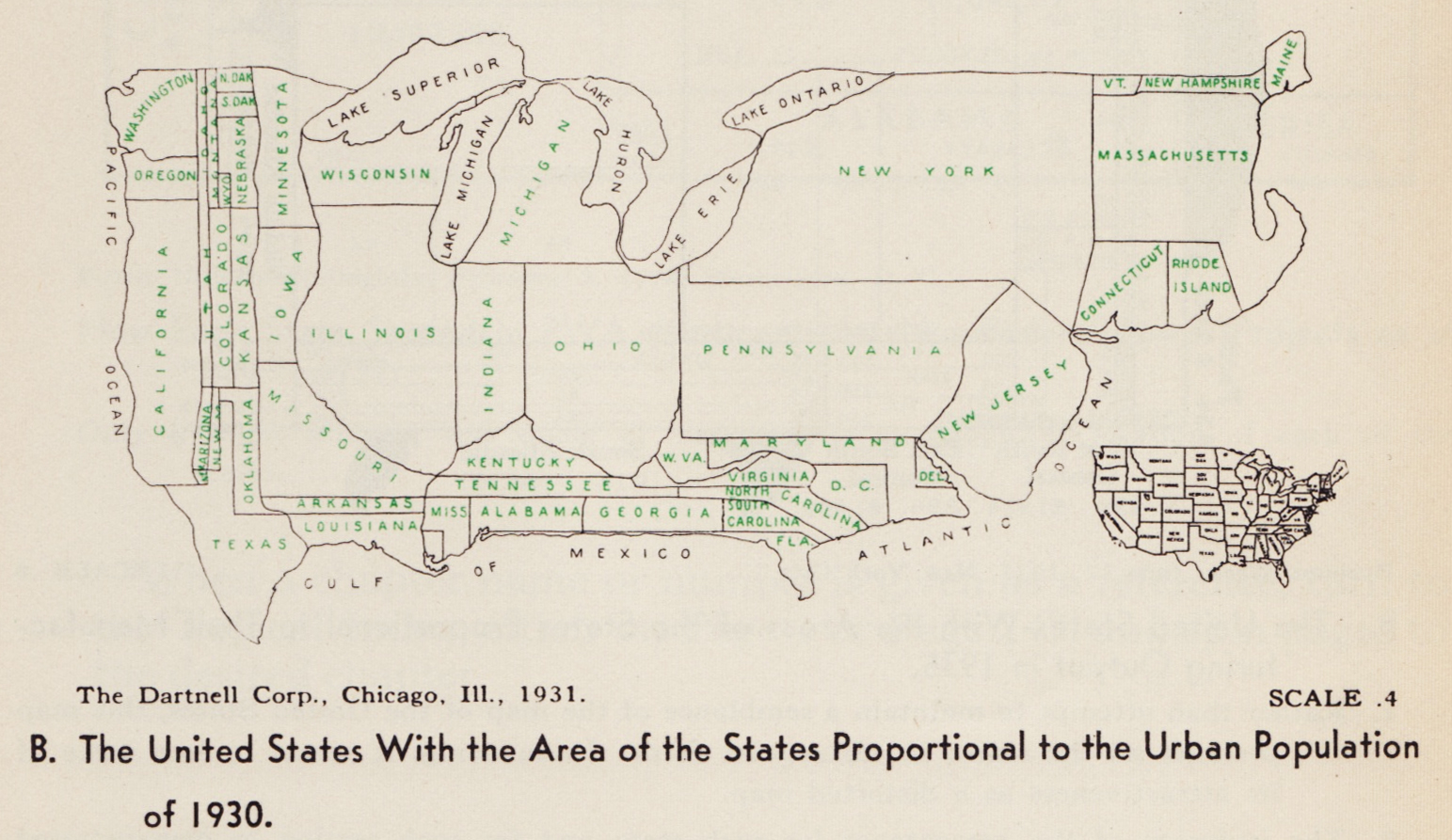}\\

(b) ``The United States With the Area of the States Proportional to the Urban Population of 1930.'' \\\vspace{.25cm}

\includegraphics[width=.49\textwidth]{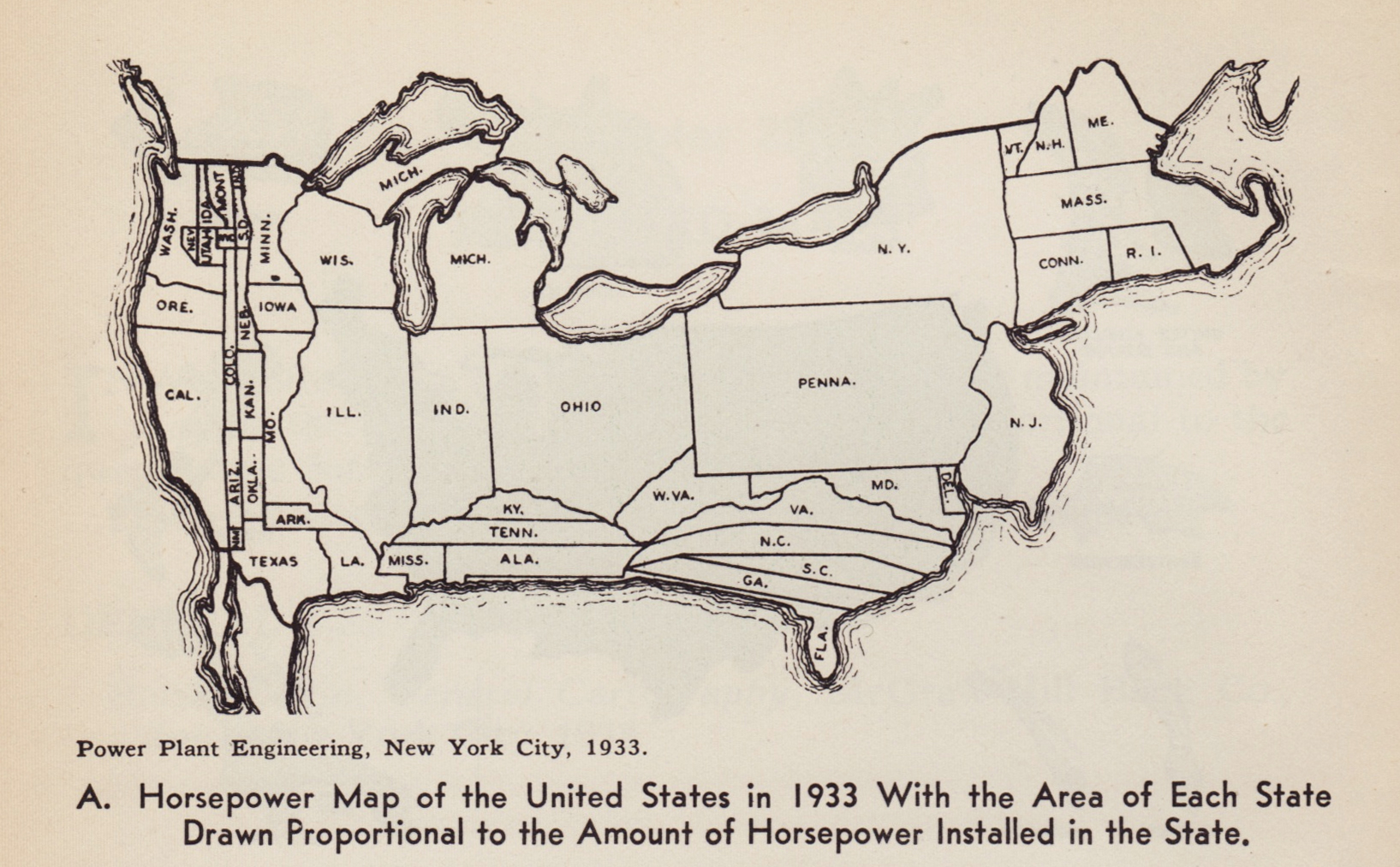}\\
(c)  ``Horsepower Map of the United States in 1933 With the Area of Each State Drawn Proportional to the Amount of Horsepower Installed in the State.'' \\
 \caption{1930s cartograms of the US from~\cite{john_carto}.}
\label{fig:cartograms_old_school}
\end{figure}

 Raisz also emphasized the educational role of a cartogram: ``Its educational value is not limited to the schools: it may serve to set right common misconceptions held by even well informed people.'' Magazines and newspapers in the United States illustrated stories with cartograms~\cite{john_carto}, such as the 1921 cartogram from {\em Literary Digest}, showing energy consumption in the US; see Fig.~\ref{fig:businessweek}(a).
{\em Business Week} used a rectangular cartogram, along with a geographical map in the background, to illustrate manufacturing output; see Fig.~\ref{fig:businessweek}(b). 
{\em Public Utilities Fortnightly} published a cartogram illustrating ``How Each State Shared in PWA Allotments for Non-Federal Power Projects as of July 1, 1937''~\cite{john_carto}; see Fig.~\ref{fig:businessweek}(c).
Figure~\ref{fig:cartograms_old_school}  shows further examples of early cartograms published between 1930 and 1938, two of which depict population while the third depicts energy production.

\begin{figure*}[t]
\begin{center}
\includegraphics[width=0.9\textwidth]{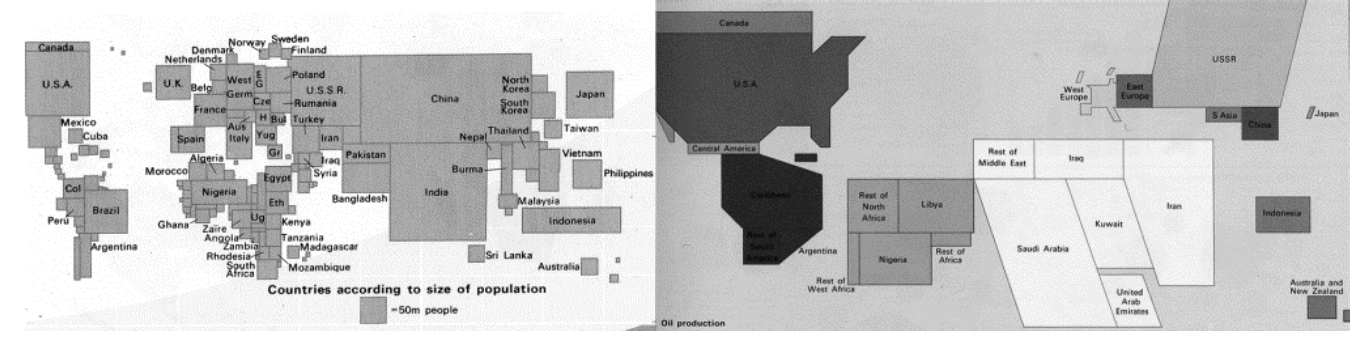}\\
\caption{Cartograms from the Atlas of Canada and the World (1979) showing (left) population and (right) oil production~\cite{school_cartography}.}
\label{fig:oil_production}
\end{center}
\end{figure*}

Cartograms have continued to be prevalent.
Figure~\ref{fig:oil_production} shows population cartograms and oil production cartograms of the world, published in the 1979 Atlas of Canada and the World~\cite{Philip_atlas}. 
Several European cartographic firms rely on cartograms to illustrate population distribution. For example, rectangular population cartograms are used in school atlases published by the German Westermann firm~\cite{German_atlas}. 
  Cartograms are also used to represent population and change in population atlases published by Oxford University Press~\cite{Oxford_atlas}. Cartograms are frequently discussed in cartography textbooks~\cite{Dent, Slocum}.
 
\section{Cartogram Literature Surveys}
Guseyn-Zade and Tikunov~\cite{guseyn1994numerical} compiled a review of several methods for computing cartograms, introducing several desired properties for cartograms-generation methods.  These properties include the reproducibility of the results and the recognizability of regions through the preservation of the original geographic shapes. 
Kocmoud~\cite{kocmoud1997constructing} further surveyed methods of generating cartograms and discussed additional desirable characteristics.
These characteristics include user-controls for shape preservation, prevention of region overlaps, and trade-offs between area accuracy and shape accuracy.

Tobler~\cite{Tobler04} contributed a comprehensive survey of cartograms. In addition to introducing the mathematics behind map projections, the survey covered the history of cartograms, methods for generating cartograms, and early work on contiguous cartograms.
 Tobler noted the difficulties in creating a cartogram that preserves shapes and relative locations at the same time. To mitigate the resulting recognition difficulties, he advocated for the use of a ``brushing'' technique in interactive environments: the cartogram should be shown next to an undistorted geographical map, and when pointing to a region in one of the maps, the corresponding region should be highlighted in the other map.
Finally Tobler reports on the necessity of cartogram evaluation measures, and in particular, three general factors: (i) statistical accuracy, (ii) shape-preservation, and (iii) computational efficiency. 

More than a decade has passed since Tobler's excellent survey. A great deal of progress has been made in the context of data visualization, as predicted by Tobler~\cite{Tobler04}: ``The computer construction of cartograms has progressed rapidly in the last several years. I expect that, with the increased speed and storage capabilities of future computers the next 35 years will lead to further changes in this field.''
We consider several motivating factors for the current survey of the state of the art:

\begin{itemize}

\item In the last decade more than 70 new papers on cartograms have appeared in journals and conference proceedings on information visualization, cartography and computational geometry. This warrants an attempt to reconsider all methods, to classify cartograms by design dimensions, and to analyze current trends and future directions.
\item Several new cartogram models have been proposed since the 2004 Tobler survey~\cite{Tobler04}. For example, the Gastner-Newman diffusion method~\cite{GN04} has become very popular in cartogram applications, likely due to its availability and its ability to generate cartograms with nearly zero area error, while maintaining recognizable region shapes. 
\item Scripting languages, such as JavaScript and Flash, have helped popularize Dorling cartograms~\cite{dorling96} for websites, blogs and online applications~\cite{NYT_O,Hans,Hans2,Proto}. 
\item While Tobler considered rectangular cartograms and their algorithmic generation, there have been new developments in rectangular~\cite{ks07,BSV12} and rectilinear cartograms~\cite{ourSoCG}.
\item Until recently, little work was done on analyzing the effectiveness of different cartograms, beyond some ad hoc performance analysis, typically comparing a new method against an existing one, using metrics such as cartographic error and running time. 
In the last decade more work has been done on evaluating cartogram algorithms, both by a standard set of performance measures and by subjective preference and user-studies~\cite{AKV15,kaspar2013empirical,NusratAK15}. 
\item Finally, 
cartograms are more frequently used as off-the-shelf tools, in social, political and public health applications.
\end{itemize}







\newcounter{dummy}

{
\begin{table*}[htbp]
\centering
\begin{tabular}{|c|c|c|c|c|c|}
\hline

\parbox{0.135\textwidth}{\centering \textbf{Type}} &
	\parbox{0.08\textwidth}{\centering \textbf{Statistics}} &
	\parbox{0.088\textwidth}{\centering \textbf{Contiguity}} &
	\parbox{0.1\textwidth}{\centering \textbf{Geography}} &
	\parbox{0.155\textwidth}{\centering \textbf{Topology}} &
 	\parbox{0.28\textwidth}{\centering \textbf{Example}}\\

\hline

\parbox{0.135\textwidth}{\centering Rubber map method \cite{Tobler73}
} &
	\parbox{0.08\textwidth}{\centering Not accurate} &
	\parbox{0.088\textwidth}{\centering Contiguous} &
	\parbox{0.1\textwidth}{\centering  Distorted} &
	\parbox{0.155\textwidth}{\centering Topology-preserving} &
 	\parbox{0.19\textwidth}{\vspace{0.1cm}\includegraphics[width=0.19\textwidth]{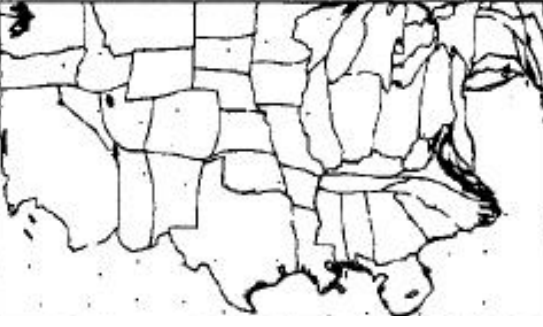}}\\




\hline

\parbox{0.135\textwidth}{\centering Pseudo-cartogram method ~\cite{tobler1986pseudo}
} &
	\parbox{0.08\textwidth}{\centering Not accurate} &
	\parbox{0.088\textwidth}{\centering Contiguous} &
	\parbox{0.1\textwidth}{\centering  Distorted} &
	\parbox{0.155\textwidth}{\centering Topology-preserving} &
 	\parbox{0.18\textwidth}{\vspace{0.1cm}\includegraphics[width=0.18\textwidth]{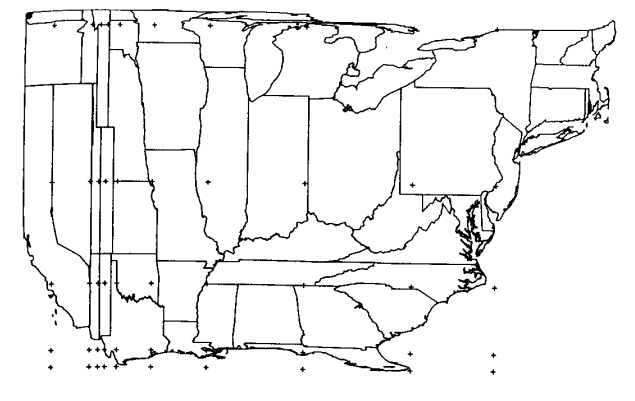}}\\

\hline


\hline



\parbox{0.135\textwidth}{\centering Constraint based approach \cite{kocmoud1997constructing}} &
	\parbox{0.08\textwidth}{\centering Not accurate} &
	\parbox{0.088\textwidth}{\centering Contiguous} &
	\parbox{0.1\textwidth}{\centering  Distorted} &
	\parbox{0.155\textwidth}{\centering Topology-preserving} &
 	\parbox{0.19\textwidth}{\vspace{0.1cm}\includegraphics[width=0.19\textwidth] {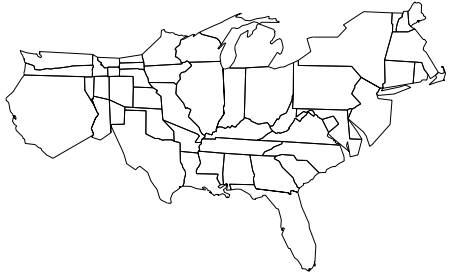}}\\

\hline

\parbox{0.135\textwidth}{\centering Cartodraw \cite{KNP04}} &
	\parbox{0.08\textwidth}{\centering Not accurate} &
	\parbox{0.088\textwidth}{\centering Contiguous} &
	\parbox{0.1\textwidth}{\centering  Distorted} &
	\parbox{0.155\textwidth}{\centering Topology-preserving} &
 	\parbox{0.19\textwidth}{\vspace{0.1cm}\includegraphics[width=0.19\textwidth]{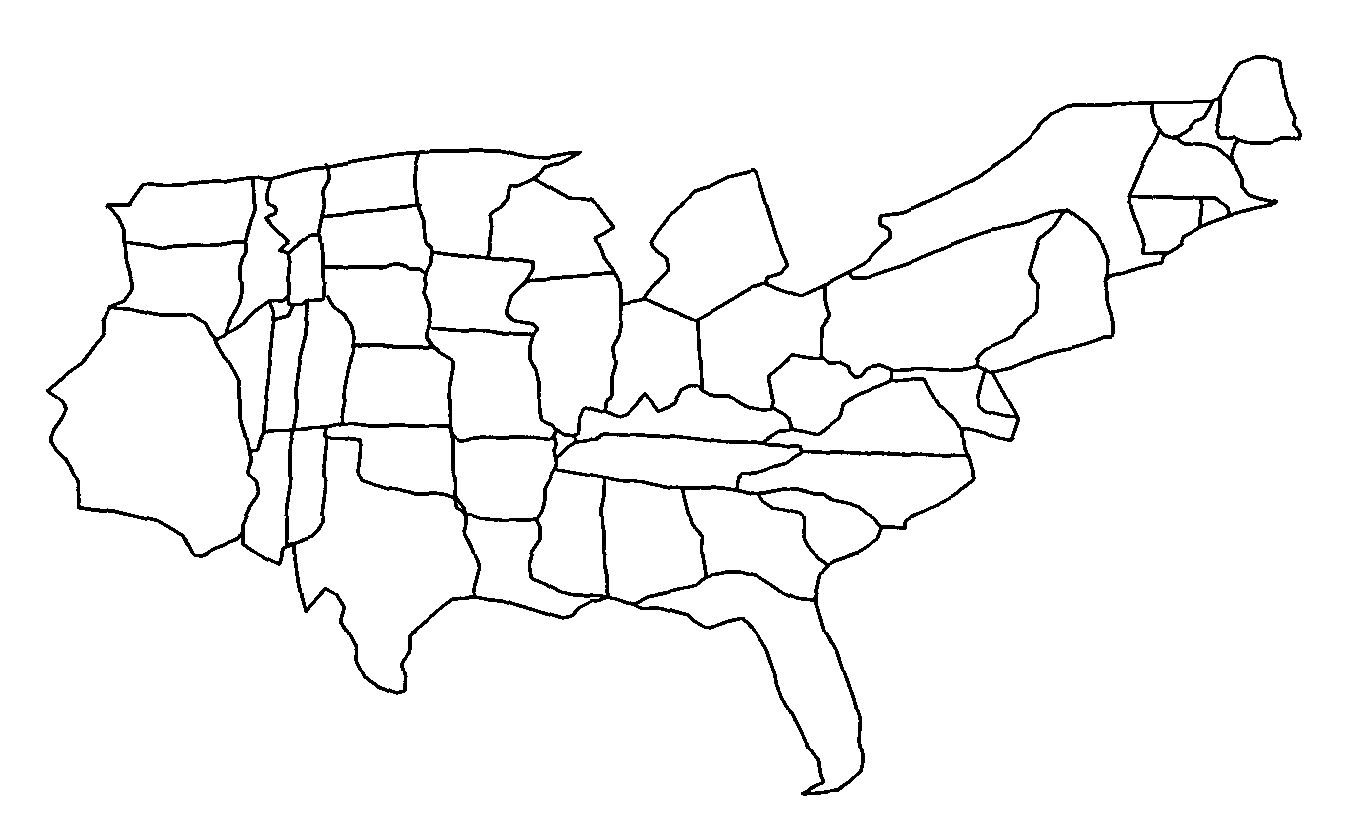}}\\

\hline

\parbox{0.135\textwidth}{\centering Medial-axis-based cartograms \cite{KPN05}
} &
	\parbox{0.08\textwidth}{\centering Not Accurate} &
	\parbox{0.088\textwidth}{\centering Contiguous} &
	\parbox{0.1\textwidth}{\centering  Distorted} &
	\parbox{0.155\textwidth}{\centering Topology-preserving} &
 	\parbox{0.19\textwidth}{\vspace{0.1cm}\includegraphics[width=0.19\textwidth,height=2.3cm]{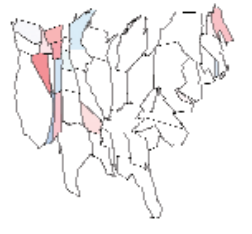}}\\

\hline



\parbox{0.135\textwidth}{\centering Cellular automata method \cite{dorling96}
} &
	\parbox{0.08\textwidth}{\centering Accurate} &
	\parbox{0.088\textwidth}{\centering Contiguous} &
	\parbox{0.1\textwidth}{\centering  Distorted} &
	\parbox{0.155\textwidth}{\centering Topology-preserving} &
 	\parbox{0.28\textwidth}{\vspace{0.1cm}\includegraphics[width=0.28\textwidth]{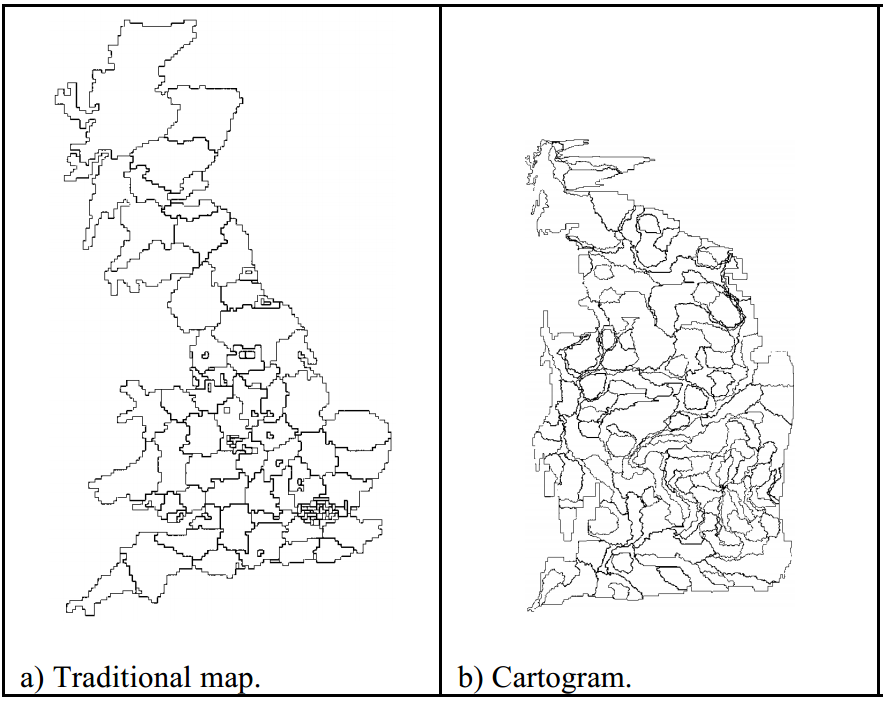}}\\

\hline

\parbox{0.135\textwidth}{\hspace{-0.1cm}\parbox{0.145\textwidth}{\centering Dorling cartograms \cite{dorling96}
}} &
	\parbox{0.08\textwidth}{\centering Accurate} &
	\parbox{0.088\textwidth}{\centering Not\\contiguous} &
	\parbox{0.1\textwidth}{\centering  Shape not preserved (circles)} &
	\parbox{0.155\textwidth}{\centering Topology not preserved} &
 	\parbox{0.19\textwidth}{\vspace{0.1cm}\includegraphics[width=0.14\textwidth]{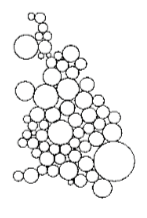}}\\

\hline

\parbox{0.135\textwidth}{\centering Rectangular cartograms \cite{BSV12} } &
	\parbox{0.08\textwidth}{\centering Depends on the variant} &
	\parbox{0.088\textwidth}{\centering Contiguous} &
	\parbox{0.1\textwidth}{\centering Shape not preserved (rectangles)} &
	\parbox{0.155\textwidth}{\centering Depends on the variant} &
	\parbox{0.18\textwidth}{\vspace{0.2cm}\includegraphics[width=0.19\textwidth,height=1.5cm]{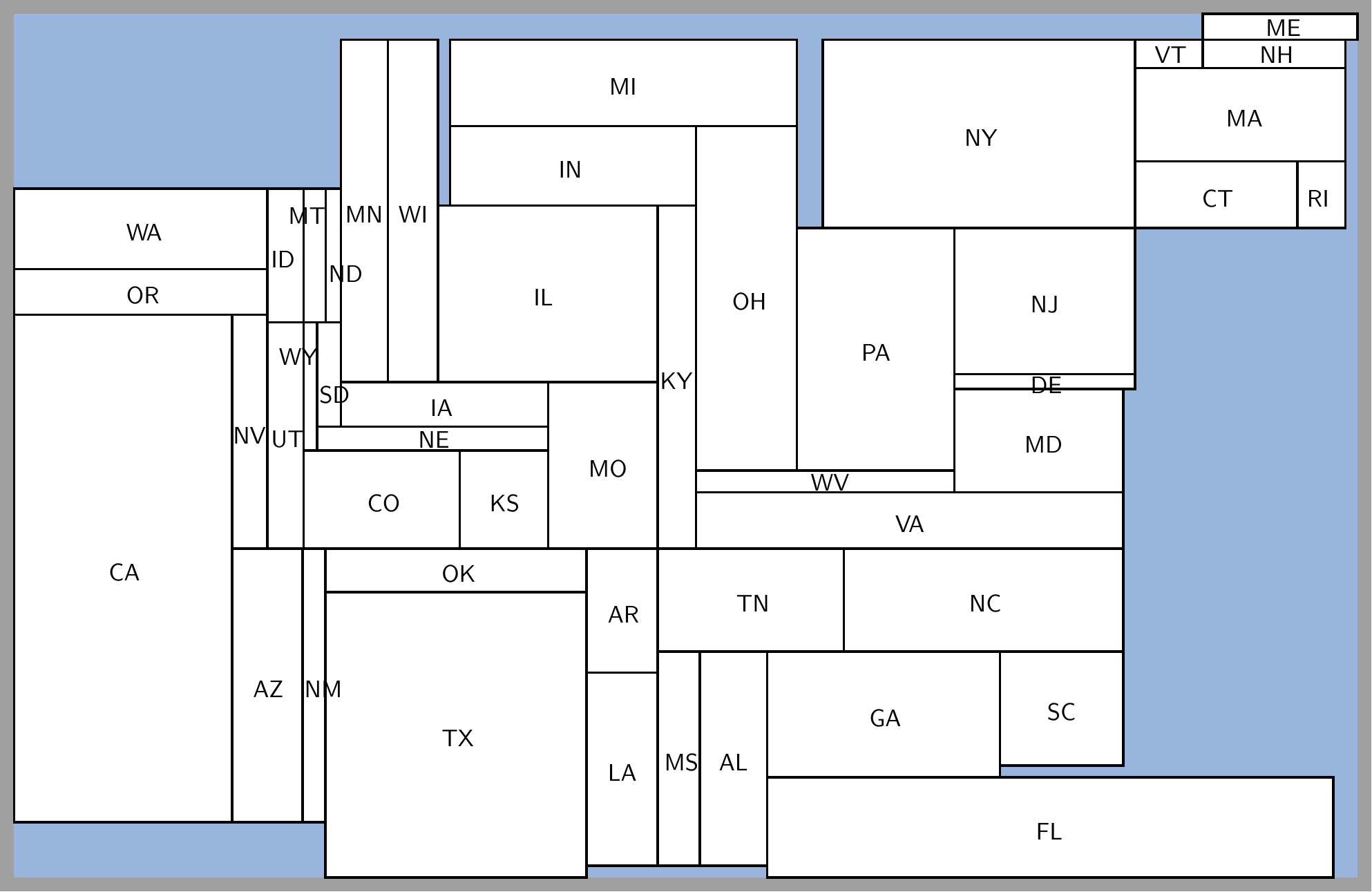} \vspace{.1cm}}\\

\hline

\end{tabular}
\caption{
An overview of
several different types of cartograms created from US and UK maps.
Images are reproduced with permission from the authors.
}
\label{table:all-algorithms}
\setcounter{dummy}{\value{table}}
\end{table*}

\begin{table*}[htbp]
\centering
\begin{tabular}{|c|c|c|c|c|c|}
\hline

\parbox{0.135\textwidth}{\centering \textbf{Type}} &
	\parbox{0.08\textwidth}{\centering \textbf{Statistics}} &
	\parbox{0.088\textwidth}{\centering \textbf{Contiguity}} &
	\parbox{0.1\textwidth}{\centering \textbf{Geography}} &
	\parbox{0.155\textwidth}{\centering \textbf{Topology}} &
 	\parbox{0.28\textwidth}{\centering \textbf{Example}}\\

\hline

\parbox{0.135\textwidth}{\centering Diffusion-based cartograms \cite{GN04}
} &
	\parbox{0.08\textwidth}{\centering Almost accurate} &
	\parbox{0.088\textwidth}{\centering Contiguous} &
	\parbox{0.1\textwidth}{\centering  Distorted} &
	\parbox{0.155\textwidth}{\centering Topology-preserving} &
 	\parbox{0.19\textwidth}{\vspace{0.1cm}\includegraphics[width=0.19\textwidth]{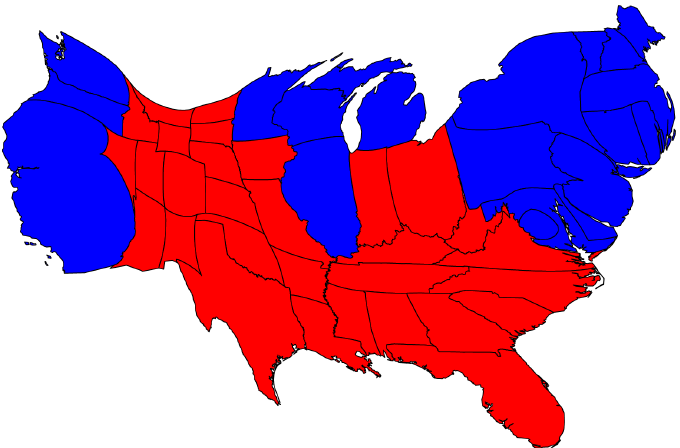}}\\

\hline

\parbox{0.135\textwidth}{\centering Circular-arc cartograms \cite{KKN13}
} &
	\parbox{0.08\textwidth}{\centering Not Accurate} &
	\parbox{0.088\textwidth}{\centering Contiguous} &
	\parbox{0.1\textwidth}{\centering  Shape mostly preserved} &
	\parbox{0.15\textwidth}{\centering Topology-preserving} &
 	\parbox{0.19\textwidth}{\vspace{0.1cm}\includegraphics[width=0.19\textwidth]{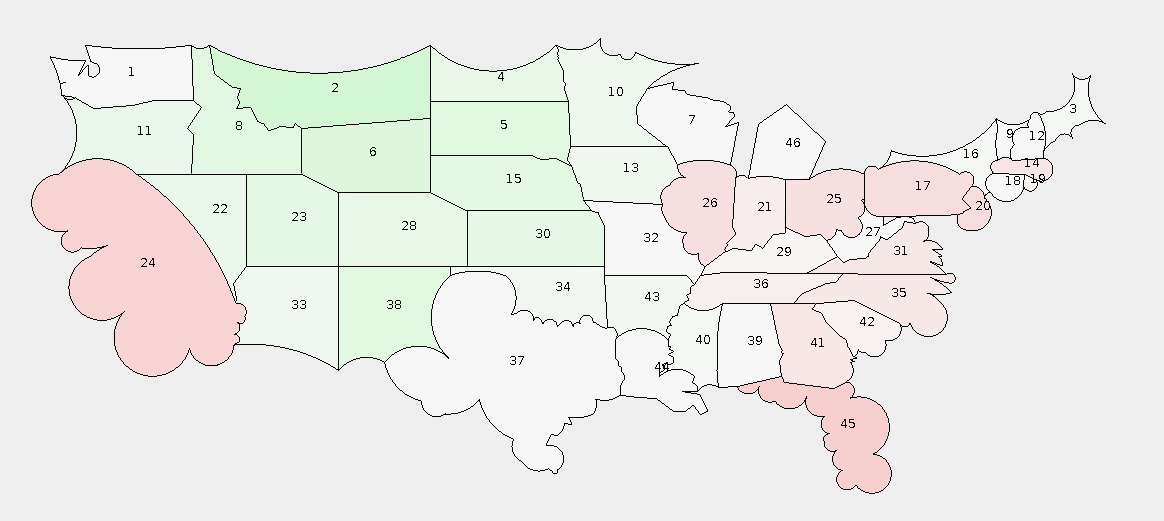}}\\

\hline

\parbox{0.135\textwidth}{\centering Optimal rubber sheet method \cite{sun2013optimized}
} &
	\parbox{0.08\textwidth}{\centering Almost accurate} &
	\parbox{0.088\textwidth}{\centering Contiguous} &
	\parbox{0.1\textwidth}{\centering  Distorted} &
	\parbox{0.155\textwidth}{\centering Topology-preserving} &
 	\parbox{0.19\textwidth}{\vspace{0.1cm}\includegraphics[width=0.19\textwidth]{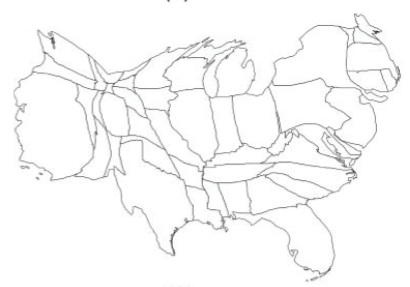}}\\

\hline

\parbox{0.135\textwidth}{\centering Fast, free-form rubber-sheet method \cite{sun2013fast}
} &
	\parbox{0.08\textwidth}{\centering Almost accurate} &
	\parbox{0.088\textwidth}{\centering Contiguous} &
	\parbox{0.1\textwidth}{\centering  Distorted} &
	\parbox{0.155\textwidth}{\centering  Topology-preserving} &
 	\parbox{0.19\textwidth}{\vspace{0.1cm}\includegraphics[width=0.19\textwidth]{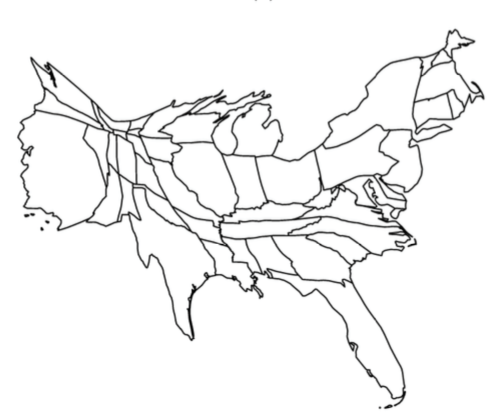}}\\

\hline

\parbox{0.135\textwidth}{\centering T-shape cartograms \cite{ourSoCG}
} &
	\parbox{0.08\textwidth}{\centering Accurate} &
	\parbox{0.088\textwidth}{\centering Contiguous} &
	\parbox{0.1\textwidth}{\centering  Shape not preserved} &
	\parbox{0.155\textwidth}{\centering Topology-preserving} &
 	\parbox{0.19\textwidth}{\vspace{0.1cm}\includegraphics[width=0.19\textwidth]{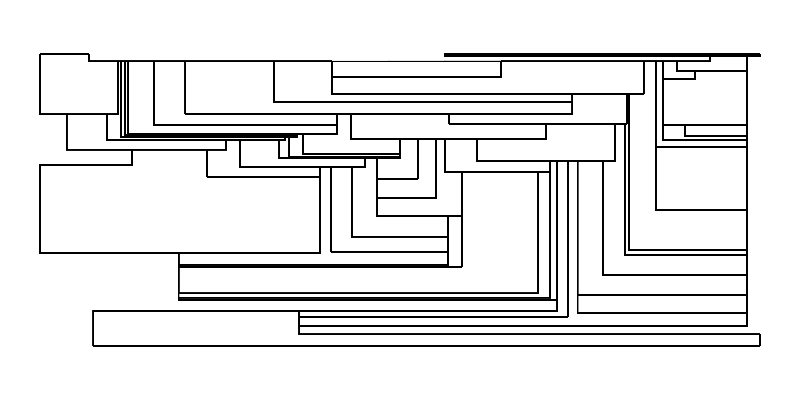}}\\

\hline

\parbox{0.135\textwidth}{\centering Non-contiguous cartograms \cite{Olson}} &
	\parbox{0.08\textwidth}{\centering Accurate} &
	\parbox{0.088\textwidth}{\centering Not\\contiguous} &
	\parbox{0.1\textwidth}{\centering Shape preserved} &
	\parbox{0.155\textwidth}{\hspace{-0.1cm}\parbox{0.165\textwidth}{\centering Topology not preserved}} &
	\parbox{0.18\textwidth}{\vspace{0.1cm}\includegraphics[width=0.19\textwidth]{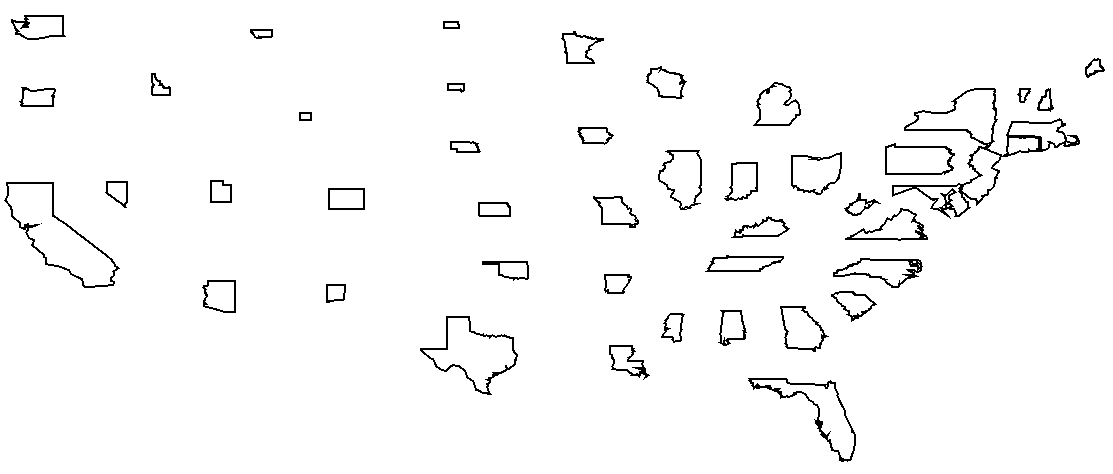}}\\

\hline

\parbox{0.135\textwidth}{\centering Demers cartograms \cite{Demers} (figure from \cite{NYT2012-elec})} &
	\parbox{0.08\textwidth}{\centering Accurate} &
	\parbox{0.088\textwidth}{\centering Not\\contiguous} &
	\parbox{0.1\textwidth}{\centering  Shape not preserved (squares)} &
	\parbox{0.155\textwidth}{\centering Topology not preserved} &
 	\parbox{0.19\textwidth}{\vspace{0.1cm}\includegraphics[width=0.19\textwidth]{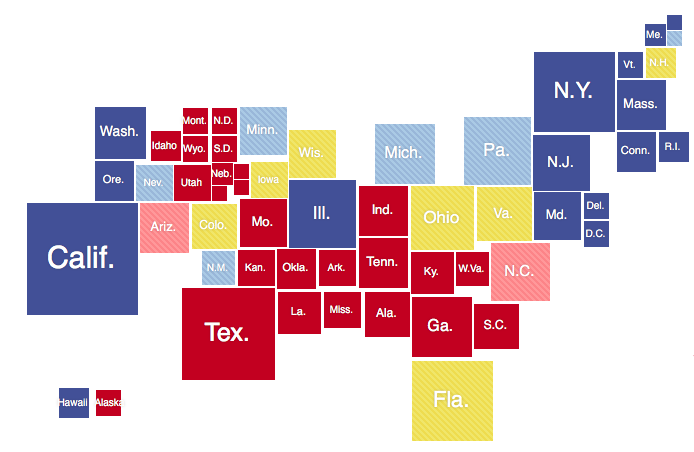}}\\

\hline

\parbox{0.135\textwidth}{\centering Mosaic cartograms \cite{cano2015mosaic}
} &
	\parbox{0.08\textwidth}{\centering Not accurate} &
	\parbox{0.088\textwidth}{\centering Contiguous} &
	\parbox{0.1\textwidth}{\centering  Shape mostly preserved} &
	\parbox{0.155\textwidth}{\centering Topology-preserving} &
 	\parbox{0.22\textwidth}{\vspace{0.1cm}\includegraphics[width=0.19\textwidth]{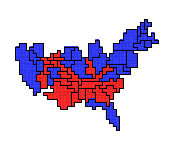}}\\

\hline

\parbox{0.135\textwidth}{\centering Table cartograms \cite{EFKKMNV13}
} &
	\parbox{0.08\textwidth}{\centering Accurate} &
	\parbox{0.088\textwidth}{\centering Contiguous} &
	\parbox{0.1\textwidth}{\centering  Shape not preserved} &
	\parbox{0.155\textwidth}{\centering Topology not preserved} &
 	\parbox{0.19\textwidth}{\vspace{0.1cm}\includegraphics[width=0.19\textwidth,height=1.9cm]{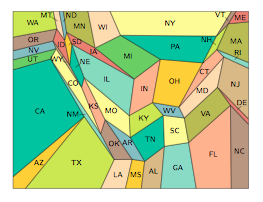}}\\

\hline

\end{tabular}
\vskip 6pt
\parbox{0.98\textwidth}
{
\textbf{Table~\thedummy~(cont.):} \textit{
An overview of
several different types of cartograms created from US and UK maps.
Images are reproduced with permission from the authors.
}
}
\vskip 6pt
\end{table*}

}


\section{Design Dimensions and Cartogram Types}

There are three major design dimensions along which cartograms may vary: 

\begin{itemize}
\item \textbf{Statistical accuracy}: Statistical accuracy refers to how well the modified areas represent the corresponding statistic shown (e.g., population or GDP). \textit{Cartographic error} measures the relative distortion of the area of each modified region from the desired statistic. Minimizing cartographic error is one of the main goals in many cartogram-generation algorithms.

\item \textbf{Geographical accuracy}: Geographical accuracy refers to how much the modified shapes and locations of the 
regions (e.g., countries and states) resemble those in the original map. {\em Shape and relative position preservation} can be measured by various metrics, such as curve similarity and pairwise distances. Preserving geographical accuracy is an explicit or implicit goal in many cartogram-generation algorithms.

\item \textbf{Topological accuracy}: Topological accuracy refers to how well the topology of the cartogram matches the topology of the original map. Perfect topology preservation is obtained when two regions are neighbors in the cartogram if and only if they are neighbors in the original map. 
Some cartogram algorithms guarantee that the topology is preserved (e.g., diffusion cartograms), while others do not (e.g., Dorling cartograms). The same terminology is sometimes applied to describe other issues, such as overlapping regions and self-intersecting borders, but in this paper we use topological accuracy in the context of correct or incorrect adjacencies between regions.

\end{itemize}

There is a wide variety of cartogram-generation algorithms, each with its own advantages and disadvantages; see Table~\ref{table:all-algorithms}.
It is worth noting that there is no ``perfect'' cartogram, that is, a cartogram that preserves shapes, preserves the topology, and has zero cartographic error. Some cartograms guarantee statistical accuracy at the expense of shape and topology distortions; others guarantee topological accuracy at the expense of shape distortions.
In addition to the three major dimensions above, another possible consideration is the {\em computational efficiency} of the cartogram-generation method. In applications involving real-time computations, as well as for interactive cartograms, the running time of the underling algorithm is also an important factor.



There are several ways to systematically categorize the large number of different types of cartograms. 
One such categorization, based on the design dimensions~\cite{ks07}, uses four major types: \textbf{contiguous, non-contiguous, Dorling, rectangular}.


\subsection{Contiguous Cartograms}
These cartograms deform the regions of a map, so that the desired size/area is obtained, while
 adjacencies are maintained. 
They are also called \textit{continuous cartograms}, or \textit{deformation cartograms}~\cite{AKV15}, since the original geographical map is modified (by pulling, pushing, and stretching the boundaries) to change the areas of the countries on the map. 
For consistency, in the rest of the paper, we refer to the cartograms of this type as {\em contiguous cartograms}.  Contiguous cartograms are topologically accurate. There is typically a trade-off between high statistical accuracy and high geographical accuracy.


Initially cartograms were created by hand and, as with many early endeavors in map-making, creating cartograms was as much art as science. Early efforts to automate this process is the Rubber Map cartogram by Tobler~\cite{Tobler73}. 
The algorithm first divides the map into a regular lattice grid and computes a density value for each grid quadrilateral. For each grid vertex, a displacement direction is computed, so that it minimizes the density errors of the four adjacent quadrilaterals. The grid vertices are moved in their displacement directions and this process is repeated until little or no improvement can be made. 
This is the first automated method for cartogram-generation and it has several potential shortcomings:
it does not converge quickly, the statistical inaccuracy can be high~\cite{kocmoud1997constructing}, and the map topology may be distorted by overlapping regions~\cite{GN04}.


A geographical map can be modified locally by focusing on points of interest, as discussed in 1978 by Kadmon and Shlomi~\cite{kadmon1978polyfocal}. Their ``polyfocal projection'' can enlarge areas of interest with a magnification function controlled by two parameters: the magnification factor of the point of focus, and the rate of change of magnification with distance from the point of focus.
A similar technique is used in the ``magnifying glass'' azimuthal map projections by Snyder~\cite{snyder1987magnifying}. 
Such map projections can be used to create cartograms, as in the Density Equalized Map Projection (DEMP) algorithm
by Selvin et al.~\cite{selvin1988transformations}. In this method, a magnification factor is computed for each region. In each iteration, one region is selected. The selected
   region changes its size to reflect the magnification factor, while its
   shape is preserved. The surrounding regions deform their shapes,
   while their areas are unchanged. This process is repeated until the areas for the regions match the desired areas. While the algorithm can achieve low cartographic error, the final cartogram heavily depends on the traversal order of the regions and the algorithm is computationally intensive~\cite{selvin1988transformations}

In order to improve the statistical accuracy and speed of computation for contiguous cartograms, Dougenik et al.~\cite{DCN85} proposed the Rubber Sheet Distortion Method. In this force-based method, each region exerts a radial force upon all the map vertices (which define the borders). This force is proportional to the region-area error and inversely proportional to distance. Another force is applied to interior vertices near a region's \textit{centroid} (the point corresponding to the arithmetic mean of all the points in the region) to prevent instability. The vertices are displaced based on a combination of these forces. This is an iterative refinement process,  which continues until the desired equal density map is obtained.
 This algorithm converges faster than Tobler's Rubber Map algorithm and achieves low cartographic error, although 
overlapping regions are still possible.

 Tobler introduced the pseudo-cartogram method~\cite{tobler1986pseudo} that provides a convenient starting point for the iterative Rubber Map algorithm. This method is designed to pre-process a map prior to cartogram construction.
Tobler considers this a ``pseudo'' cartogram because it only provides an approximation of a contiguous 
cartogram. Instead of enlarging or shrinking the regions themselves, the algorithm moves the region's associations to a reference grid (similar to latitude-longitude), in order to achieve the same effect. This cartogram maintains the relative positions (north-south or east-west) between regions quite well, but it often contains extensive cartographic error~\cite{Demers}.

Torguson~\cite{torguson1990cartogram} introduced an interactive ``polygon zipping'' method to create cartograms. For this cartogram, the viewer scales and rotates each region independently along the $x$ and $y$ axes to achieve desired areas. The viewer then positions the resulting regions as close as possible. Adjacent region edges are then ``zipped'' together by an edge matching algorithm.
 This method has some drawbacks: the viewer might incorrectly ``zip'' distant regions, and the statistical and topological accuracy is highly dependent on the effort and skill of the viewer.

 Another approach to generate contiguous cartograms comes from Dorling's cellular automata method~\cite{dorling96}. The input map is first divided into checkerboard grids. Then the regions with more than required number of cells will pass them to those that have fewer than required, until the desired number of cells is reached for each region. The procedure is elegant and simple, although the resulting cartograms do not preserve shapes.

Kocmoud~\cite{kocmoud1997constructing} proposed a constraint-based approach to generate contiguous cartograms where two distinct and conflicting tasks are considered: achieving desired areas without regard to shape, and then restoring shape while attempting to keep the areas fixed. The algorithm modifies the map incrementally, applying these two steps until a desired accuracy is obtained. This algorithm makes it possible to control the balance between statistical accuracy and shape accuracy. However, the algorithm
converges rather slowly~\cite{GN04} and can create significant deformation of the global shape~\cite{KNP04}. 

The \textit{CartoDraw} algorithm proposed by Keim et al.~\cite{KNP04} deforms a map into a cartogram using a set of ``scanlines'', that are either computed automatically or determined manually.
A distance from the original map to the cartogram is defined with a metric based on a Fourier transformation. The scanline algorithm repositions the edges according to this metric. Each map vertex within a certain distance from a scanline is considered for repositioning, and is indeed repositioned if the resulting area error and shape distortion are within a given threshold. This algorithm is relatively fast, but its increased speed depends on schematizing the input map by polygons with fewer vertices, and might have non-trivial cartographic error~\cite{GN04}. 

In the medial-axis algorithm~\cite{KPN05}, Keim et al. proposed a similar algorithm to CartoDraw, where the medial axes of the polygonal regions are used as the scanlines. VisualPoints by Keim et al.~\cite{KNPS03} created cartograms, based on quadtree partitions of the plane. Homeomorphic deformations are used to create the contiguous cartograms of Welzl et al.~\cite{edelsbrunner1995combinatorial}.  

In 2004, Gastner and Newman proposed a new diffusion method to generate contiguous cartograms~\cite{GN04}. In this method, the original input map is projected onto a distorted grid, computed in such a way that the areas of the countries match the pre-defined values. They express the problem as an iterative diffusion process, where quantities flow from one country to another until a balanced distribution is reached, i.e. the density is the same everywhere. 
This method allows for minimal cartographic error, while also keeping region shapes relatively recognizable. Over the last decade, this has become one of the most popular techniques to create cartograms. Its popularity is likely due to its shape recognizability, and the availability of the software~\cite{carto_sw} to generate these cartograms. 

Inoue and Shimizu~\cite{inoue2006new} proposed an algorithm that aims to provide a trade-off between statistical accuracy and geographical accuracy.
Regions in the input map are divided into triangles using a Delaunay triangulation. The statistical data for each region are distributed among its triangles in proportion to their sizes. The triangles in the map are then modified to realize the assigned data, subject to a constraint in the changes of the angles between adjacent edges in the triangles. This ``regularizing condition'' attempts to preserve the shape of the regions.  
The trade-off between statistical accuracy and shape change is parameterized with a single user-controlled parameter. 
The method is computationally efficient and can result in high statistical accuracy.
 However, the input map and the resulting cartogram is often highly schematized. More triangles are needed to create less schematized cartograms, which results in increased computation time.

The contiguous cartogram-generation algorithm of Sagar~\cite{sagar2013cartograms} is based on {\em mathematical morphology}. The algorithm first computes the (morphological) centroid of each region by iteratively peeling the polygon until only a single point (the \textit{centroid}) remains. It then simulates a simultaneous flood propagation from the centroids for each region. This is done by considering circles of iteratively increasing radius, centered at the centroid of each region. The flood propagation speed (i.e., the rate of growth of the radius) for a region depends on value of the statistical variable for that region. By maintaining the envelope of these flood propagation around each region, a modified shape for the region is obtained.
The resulting cartogram preserves the overall shape of the map (i.e., the outer boundaries), but distorts the shapes of the individual regions and has non-trivial cartographic error.


A new variant of the rubber-sheet algorithm by Sun~\cite{sun2013optimized} attempts to to improve on some of the problems associated with this class of algorithms (e.g., overlapping regions and high computational cost). The new variant converges faster and has fewer overlapping regions. Yet another variant, \textit{Carto3F}~\cite{sun2013fast}, uses an auxiliary quadtree data structure, simplified polygon shapes, and allows for parallelized force computations to speed-up the cartogram-generation. This variant also avoids overlapping regions.


Despite the great variety of contiguous cartogram-generation methods, it seems that the Gastner and Newman diffusion method~\cite{GN04} has been the most popular in the last decade.  
Tools to generate diffusion cartograms are available online~\cite{scapetoad}, the underlying algorithm is computationally efficient, and the resulting maps tend to provide an acceptable balance between statistical and geographical accuracy. 
Worldmapper~\cite{WorldMapper} maintains a good collection 
of diffusion-based cartograms.

\subsection{Non-Contiguous Cartograms}
These cartograms are created by starting with an undistorted geographical map and scaling down each region independently, so that the desired areas are obtained. 
The result is a piece-wise contiguous but overall non-contiguous cartogram~\cite{Olson}. Non-contiguous cartograms provide perfect statistical accuracy and perfect shape preservation. However, they fail to preserve the topology of the original map.
These cartograms are easy to generate and there is some evidence that loss of the topology of the original map might not cause major perceptual problems~\cite{KPN05}. Olson~\cite{Olson} lists three useful properties of non-contiguous cartograms: 
\begin{itemize}
\item The gaps between the enumeration units (e.g., states) denote the discrepancies of values between them.
\item The representation and manipulation of this cartogram involves only resizing the discrete units.
\item The shapes and positions of the regions do not change, only their relative sizes do. Therefore, recognition of units is less complicated.
\end{itemize}

However, since the regions do not touch each other, the topology of the map is difficult to comprehend. Furthermore, since the sizes of the regions depends on their original size and statistic to be shown, some regions may become too small. Mapping Worlds~\cite{non_c} uses non-contiguous cartograms to illustrate different social and political data with interactive features.

\subsection{Dorling Cartograms}
This type of cartogram was named after its creator, Danny Dorling~\cite{dorling96}. In this cartogram, regions are represented by circles. The statistic of interest is realized by the sizes of the circles: the bigger the circle, the larger the data value. Two types of forces work on these circles to avoid overlap: a repulsive force pushes overlapping circles away, and an attraction force tries to keep the circles close to their initial positions. 
An iterative force-directed movement improves the original configuration until overlaps are eliminated. 
Unlike contiguous and non-contiguous cartograms, Dorling cartograms preserve neither shape nor topology. Nonetheless, they can guarantee zero cartographic error, and are popular, due to
 JavaScript libraries such as D3~\cite{bostock2011d3, D3} and Protovis~\cite{Proto}. 
 

A Demers cartogram~\cite{Demers} is a variant of a Dorling cartogram, where squares are used in place of circles.  
Demers cartograms have no cartographic errors, but do not preserve shapes. Since squares can be packed more compactly than circles, Demers cartograms can capture the underlying map topology better than Dorling cartograms. In a way, Demers cartograms can be thought of as a special case of rectangular cartograms.

\subsection{Rectangular Cartograms}
These cartograms represent regions by rectangles. The size of each rectangle corresponds to the variable of interest. Similar to Dorling and Demers cartograms, rectangular cartograms do not preserve shapes. Some variants make it possible to achieve zero cartographic error, but at the expense of topological errors. In general, rectangular cartograms cannot guarantee topological accuracy. For example, consider four countries, each of which are pairwise neighbors: the corresponding dual graph of such a map is the complete graph $K_4$, and this graph cannot be represented by contact of rectangles. 

Rectangular cartograms have been used for more than 80 years~\cite{Raisz34}. 
The first automated method to produce rectangular cartograms is
 \textit{RecMap} by Heilmann et al.~\cite{hkps04}. Here the map is represented by a partition of a rectangle into smaller rectangular shapes realizing the map regions. The areas of these rectangular regions are proportional to the given statistical values. The rectangles are placed as close as possible to their original positions and as close as possible to their neighbors. Heilmann et al. considered two variants: one gives zero empty space, and the other attempts to minimize shape distortion, while both variants maintain zero cartographic error. 
 The cartograms are generated by a genetic algorithm which evaluates generations of ``candidate cartograms'' with an objective function combining statistical, geographical and topological accuracy. The most fit candidates are selected to produce the next generation, via replication and mutation. However, by the nature of the algorithm, topological accuracy and geographical accuracy cannot be guaranteed.
 

The algorithms by van Kreveld and Speckmann~\cite{KS05,ks07} improve on the rectangulation method. An initial rectangular layout is first computed from the topology of the map. A segment-moving heuristic is then used to compute appropriate coordinates for the rectangles, so that they realize the desired areas. In this heuristic, horizontal and vertical segments of the rectangular layout are iteratively moved to reduce the cartographic error. This can be applied in two different settings: either the topology might be disturbed to achieve nearly perfect area, or the topology is preserved at the cost of cartographic errors. This simple segment moving heuristic was later improved in~\cite{SKF06} with a linear programming approach.

There can be many rectangular layouts for the same map. The evolution algorithm in~\cite{BSV12} uses this fact to design an improvement from the simple segment moving heuristic. The algorithm finds the ``fittest'' rectangular cartogram for a map using a genetic algorithm strategy. At each step the algorithm takes a number of different rectangular layouts for the map and keeps only those with the best scores. Here the score for a cartogram is a function that combines the cartographic error and distortion of the cartogram from the map. New rectangular layouts are generated from recombining the fittest old ones.

In addition to cartographic errors or topological errors, rectangular cartograms have another potential problem. To make a map realizable with a rectangular cartogram, it might be necessary to merge two countries into one, or split one country into two parts. For example, in a cartogram of Europe, the region representing Luxembourg either gets merged with one of its neighboring countries or one of its neighbors gets split into two parts~\cite{ks07}. In both cases, to show the correct countries, at least one country will no longer be represented by a rectangle, but by a rectilinear region. Note that this problem is related to the impossibility of realizing the complete graph$K_4$ by a contact of rectangles.

\subsection{Other Notable Cartogram Variants}


Since Tobler's 2004 survey, there have been several new cartogram methods, such as circular-arc cartograms, rectilinear cartograms, table cartograms, and mosaic cartograms. 

In circular-arc cartograms by K\"amper et al.~\cite{KKN13}, the straight-line segments of a map are replaced by circular arcs. The curvature of the circular arcs can be used to ``inflate'' the regions with less area than required and ``deflate'' those with more area than required. In the resulting circular-arc cartograms, the regions are of two types --  ``clouds'' and ``snow-flakes'' -- making it easy to determine whether a region has grown or shrunk, just by observing its overall shape. 



\textit{Rectilinear cartograms} are a generalization of rectangular cartograms, where the requirement of rectangular shapes is relaxed by allowing regions to be formed by axis-aligned (rectilinear) polygons. These cartograms are generally topological, in the sense that they take the topology of the map into account and generate a rectilinear schematization with prescribed areas for the regions. Thus these cartograms can guarantee statistical accuracy (zero cartographic error) and  topological accuracy (all country adjacencies preserved). Note however, that these cartograms do not preserve the shapes or relative positions of the regions in the map. Rectilinear cartogram have been studied for over a decade, with the main goal of reducing the polygonal complexity (number of sides per region) from the initial 40~\cite{deBerg07} to 34~\cite{Nagamochi}, then to 12~\cite{ElenaTR}, 10~\cite{ourISAAC}, and finally to 8~\cite{ourSoCG}. Another algorithm for rectilinear cartograms modifies an initial rectangular cartogram into a rectilinear one with no cartographic error, while attempting to use as few sides per regions as possible~\cite{BMS10}.


Evans et al.~\cite{EFKKMNV13} designed and implemented, what they refer to as, ``table cartograms.'' The input is a two dimensional (grid-like) table with non-negative weights for each grid cell such that the total area equals the sum of the weights. The algorithm then modifies the grid cells into convex quadrilaterals, while maintaining the same adjacencies, realizing the weights as areas, and keeping the shape of the outside boundary fixed. In practice, the topology of a map rarely resembles a grid-like table. However, a recent grid-map layout of Eppstein et al.~\cite{EKSS15_j} offers a practical and straightforward approach to creating grid representations for geographical maps. In such a schematization, the regions of a geographical map are mapped onto a 2D square grid, in a way that preserves as much as possible the adjacencies and relative positions of the corresponding regions. Once such a mapping is computed, the table cartogram algorithm can be applied to compute a cartogram of the original map. For many countries with simple state-structure, such as France, Italy and even the USA, table cartograms can be a viable alternative to rectangular cartograms.
Table cartogram have no cartographic error, but do not preserve shapes. The extent of topology preservation depends on how similar the original map is to a grid.

Cano et al.~\cite{cano2015mosaic} designed and implemented ``mosaic cartograms.'' These cartograms redraw the input geographical map as a tiling of the plane, using simple tiles (squares or hexagons). Especially when data values are small integer units (e.g., number of congressmen in a state), this type of cartogram makes it easy to compare the statistical values of regions of interest. Mosaic cartograms can achieve low cartographic error, while maintaining correct adjacencies and recognizable shapes of the regions.


\section{Tasks}

In order to choose the most effective type of cartogram for a specific application, we need to understand the set of tasks that are to be performed. There are many task taxonomies in information visualization and cartography, as detailed below.

Wehrend~\cite{Weh} defined ``visualization goals'' as actions a user may perform on their data. Zhou and Feiner~\cite{ZHOU} specified ``visualization techniques'' as low-level operations. They also defined ``visual tasks'' as interfaces between high-level presentation intents and low-level visual techniques, without specifying exactly ``how'' an operation is done. Amar et al.~\cite{AMAR} presented a list of low-levels tasks that capture activities one engages in while using information visualization tools to help understand the given data. Yi et al.~\cite{YI07} proposed general categories of tasks used in interactive information visualization that represent ``user intents'' when interacting with a system (rather than the low-level interaction techniques provided by the system).


The typology of abstract visualization tasks proposed by Brehmer and Munzner~\cite{BM13} focused on three questions: {\em why} is a task performed, {\em what} are the inputs and outputs, and {\em how} is the task performed. Other classifications by B\"orner et al.~\cite{KATY}, Roth~\cite{RR13}, and Schulz et al.~\cite{HTMH13} add the following: 
 {\em where} in the data does a task operate? {\em when} is a task performed? {\em who} is executing a task? Overall, the collection of questions relate to the goals, the means, the characteristics, the target and cardinality of data entities, the order of the tasks, and the type (expert/non-expert) of audience.

\subsection{Task Taxonomy for Cartograms}

Although there are many task taxonomies in cartography, information visualization, and human-computer interaction, there is only one specific taxonomy for cartogram tasks~\cite{Task_C}.
Tasks are categorized in four dimensions (goals, means, characteristics and cardinality), based on four of the six questions (why, how, what, and where) from~\cite{HTMH13}.
Here we list the ten tasks in the task taxonomy, along with a general description and specific examples.

\begin{enumerate}[1.]
\item \textbf{Detect change:} 
 In cartograms the size of a country is changed in order to realize the input weights. Since change in size (i.e., whether a region has grown or shrunk) is a central feature of cartograms, the viewer should be able to detect such change.

\textit{Example Cartogram Task: Given a population cartogram of the USA, can the viewer detect if the state of California has grown or shrunk?}

\item \textbf{Locate:} The objective of this task is to search and find the location of a region in a cartogram.


\textit{Example Cartogram Task: Given a population cartogram of the USA, locate the state of California.}

\item \textbf{Recognize:} This goal of this task is to see if the shape of a region is recognizable.


\textit{Example Cartogram Task: Given the shape of a region from the original map and shapes of two regions from the cartogram, find out which of the two cartogram regions corresponds to the region from the original map.}

\item \textbf{Identify:}
For cartograms, \textit{identify} is used for attribute or characteristic search focused on a single object.

\textit{Example Cartogram Task:  If US election results are shown in a red-blue cartogram, identify the winning candidate for the state of California.}

\item \textbf{Compare:}
As the name implies, this task is used to compare; i.e., find similarities or differences between attributes.

\textit{Example Cartogram Task:  Given a population cartogram of the USA, compare two regions by size.}

\item \textbf{Find top-k:}
 The goal of this task is to find $k$ entries with the maximum
 (or minimum) values of a given attribute.

\textit{Example Cartogram Task: Given a population cartogram, find out which region has the highest/lowest population. }

\item \textbf{Filter:}
The  \textit{filter} task asks to find data that satisfies some criteria about a given attribute, 
by filtering out examples that fail the criteria.

\textit{Example Cartogram Task: Find states which have higher population than the state of California.}

\item \textbf{Find adjacency:}
The goal of this task is to identify neighbors, which can be difficult in some types of cartograms.

\textit{Example Cartogram Task:  Given a cartogram, find all adjacent states of California.}

\item \textbf{Cluster:}
The goal of this task is to find ``clusters'' or objects with similar attributes.

\textit{Example Cartogram Task:  Given a cartogram with obesity rates encoded by color, find states which have a similar obesity rate to that of California.}

\item \textbf{Summarize:} This task aims to find data patterns and trends. 
Cartograms are most often used to convey a ``big picture''. The \textit{summarize} task is one that asks the viewer to see the big picture. 

\textit{Example Cartogram Task:  Given a red-blue presidential election results cartogram, determine whether it was a close election, or a ``landslide win.''}

\end{enumerate}


\section{Evaluation of Cartograms}

Here we consider quantitative evaluation of cartograms, where various performance measures are defined and used to evaluate cartograms. We also review several task-based studies, examining the qualitative effectiveness of different cartograms. We then discuss evaluation studies that look into subjective preferences and consider interactivity in cartograms.

\subsection{Evaluation Based on Performance Measures}

\begin{figure}[h]
\centering
\includegraphics[width=0.15\textwidth]{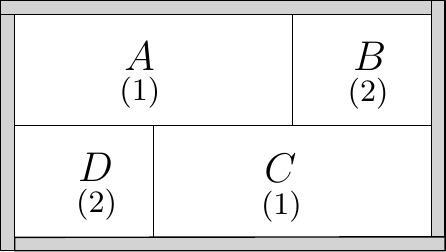}
\hspace{0.02cm}
\includegraphics[width=0.15\textwidth]{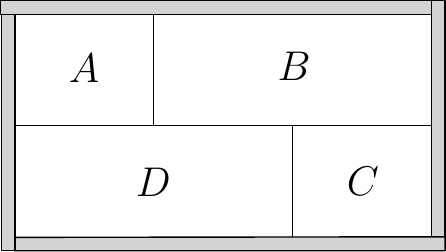}
\hspace{0.02cm}
\includegraphics[width=0.15\textwidth]{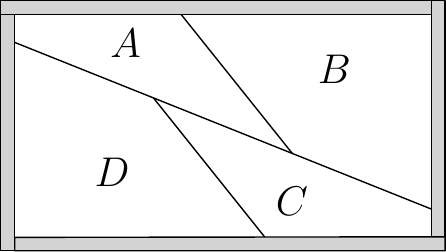}\\
(a)\hspace{0.14\textwidth}(b)\hspace{0.14\textwidth}(c)
\caption{ Three cartograms for a map with an area assignment for four states
 ($A$, $B$, $C$, $D$ with desired areas $1$, $2$, $1$, $2$, respectively),
 containing (a) cartographic error, (b) topology error and (c) shape error.
There is no cartogram with no cartographic error, topological error and shape error~\cite{AKV15}.}
\label{fig:counter}
\end{figure}

The three parameters most frequently used for evaluating cartograms are {\em cartographic error} (how accurately do modified areas match the desired data values), {\em shape error} (how similar are  the modified regions to the original regions), and {\em topological error} (how well are the adjacency relations maintained); see Fig.~\ref{fig:counter}.

Given a cartogram of a map, the \textit{individual cartographic error} for a region $v$ is defined as
 $|o(v)-w(v)|$, where $o(v)$ is the area of the region for $v$ in the cartogram and $w(v)$ is the desired statistical value for $v$~\cite{KNPS03}. This value is {\em normalized} by one of three factors:  (i) the required area $w(v)$, as in~\cite{ks07,BSV12}, (ii) the summation $o(v)+w(v)$, as in~\cite{KNP04}, and
(iii) the maximum of $o(v)$ and $w(v)$. Alam et al.~\cite{AKV15} argue for the maximum of $o(v)$ and $w(v)$ as the normalization factor, since it provides the most uniform and symmetric behavior for cartographic error. Finally, by combining the individual normalized errors for all regions, we obtain the \textit{overall cartographic error}. Thus, two standard ways to measure statistical distortions are: the \textit{average cartographic error}, $\frac{1}{|V|}\displaystyle\sum_{v\in V}\frac{|o(v)-w(v)|}{max\{o(v),w(v)\}}$, and the \textit{maximum cartographic error}, $\displaystyle\max_{v\in V}\frac{|o(v)-w(v)|}{max\{o(v),w(v)\}}$.

Several parameters have also been used to compute \textit{shape error}. Arkin et al.~\cite{ACH91} computed a translation-invariant, scale-invariant, and rotation-invariant parameter for the deviation between two polygons, by normalizing them
 by perimeter and then measuring a turning function (which captures turning angle and edge length). 
Keim et al. consider a similar measure (also translation-invariant, scaling-invariant, and
 rotation-invariant)~\cite{KNPS03}, based on a Fourier transformation of the turning angle functions.
Heilmann et al.~\cite{hkps04} use only the aspect ratios of the axis-aligned bounding boxes
 when comparing the shapes of polygons. Alam et al.~\cite{AKV15} evaluate all these parameters and suggest the use of the \textit{Hamming distance} as a measure for shape distortion. The Hamming distance~\cite{SKI98} is also known as symmetric difference~\cite{MRS10} between two polygons. Two polygons are superimposed and the fraction of area that is in exactly one of the polygons determines the Hamming distance.
 For scale-invariance, polygon areas are normalized before the measure, and for translation-invariance, the smallest error among all possible values of translation (up to a small discretization) is considered.
 
\textit{Topological error} is measured by the fraction of the adjacencies that the cartogram fails to preserve~\cite{hkps04}. The parameter is calculated as $1 - \frac{|E_c\cap E_m|}{|E_c\cup E_m|}$, where $E_c$ and $E_m$ are respectively the adjacencies between countries in the cartogram and the map.

The cartographic error, shape error, and topological error, along with occasional ad hoc measures, have been used to evaluate cartograms in different studies.
 For example, Keim et al.~\cite{KNPS03} used both cartographic error and shape error to analyze the relative performance of two algorithms: CartoDraw and VisualPoints. Buchin et al.~\cite{BSV12} also used cartographic error in the performance evaluation of rectangular cartograms.  de Berg et al.~\cite{BMS10} presented an algorithm for constructing rectilinear cartograms with zero cartographic error and correct region adjacencies. They compared their cartograms by the polygonal complexity (number of corners) and a measure of ``fatness'' of the polygonal regions used. Henriques et al.~\cite{Carto_SOM} proposed an algorithm Carto-SOM and compared it with  existing cartogram algorithms, by computing cartographic error and by visual analysis. More recently, Alam et al.~\cite{AKV15} proposed a set of quantitative measures in four dimensions (statistical distortion, topology distortion, orientation and shape distortion, and polygonal and runtime complexity) and analyzed several different types of cartograms using these measures.

\subsection{Task-based Evaluation}

Dent~\cite{dent1975} was one of the first to test the effectiveness of cartograms and wrote that ``attitudes point out that these (value-by-area) cartograms are thought to be confusing and difficult to read; at the same time they appear interesting, generalized, innovative, unusual, and having -- as opposed to lacking -- style''. 
Dent also suggested some techniques for effective communication of cartograms if the audience is not familiar with geographical shapes of statistical units, such as providing an inset map and labeling the statistical units on the cartogram.
Following Dent, Griffin~\cite{Gri83} studied the task of identifying locations in cartograms and found that cartograms are effective. Olson~\cite{Olson} designed methods for the construction of non-contiguous cartograms and studied their characteristics. 
Krauss~\cite{Krauss_ms} studied non-contiguous cartograms  as a means of communicating geographical information. She chose three distinct evaluation tasks from the range of very general to specific in order to find out how well the geographical information is communicated by cartograms, and found out that non-contiguous cartograms worked well for showing general distribution of information, but did not work well
for showing specific information (ratios between two states). Experiments involving cartograms and choropleth maps have been conducted to test if ``map labeling served as a cued recall task to determine whether theme data could be associated with administrative regions.''~\cite{rittschof1994comparing, rittschof1998learning}. The results showed that the use of feature information (value shading) in choropleth maps was more advantageous than the use of structural information (area distortion) in cartograms for depicting data to be remembered. However, these results might be influenced by the relatively simple map used.

In a more recent study, Kaspar et al.~\cite{kaspar2013empirical} investigated how people make sense of population data depicted in contiguous (value-by-area) cartograms, compared to choropleth maps which were combined with graduated circle maps. The subjects were asked to perform tasks, based on Bertin's map reading levels (\textit{elementary}, \textit{intermediate} and \textit{overall})~\cite{BERTIN83}. The overall results showed that choropleth/graduated circles are more effective (as measured by accurate responses) and more efficient (as measured by faster responses) than the cartograms.
The results seemed to depend on the complexity of the tasks (simple tasks are easier to perform in both maps compared to complex tasks), and the shapes of the polygons. Note that only one type of cartogram (Gastner-Newman diffusion cartogram) was used in this study.

A recent study~\cite{NusratAK15} evaluated four major types of cartograms (contiguous, non-contiguous, rectangular, and Dorling), based on the task taxonomy for cartograms.
The results of this study show significant differences in performance (in terms of task completion time and accuracy) between the four types of cartograms. Different tasks seem better suited to different types of cartograms. 
Achieving cartogram perfection (with respect to minimum cartographic error, shape recognizability and topology preservation) is difficult and no cartogram is equally effective in all three dimensions. Rectangular cartograms which preserve topology perform well on \textit{find adjacency} tasks. Non-contiguous cartograms maintain perfect shapes, making the \textit{recognize} task easy, but the ``sparseness'' of the map makes it difficult to understand adjacencies. Dorling cartograms disrupt the adjacency relations but somewhat preserve the relative positions of regions, and are good at the ``big picture'' \textit{summarize} task. Contiguous cartograms (specifically the Gastner and Newman diffusion variant) give the best performance for most of the tasks, although they might dramatically distort region shapes. Therefore, the choice of cartogram should depend on what needs to be shown and  what visualization tasks are the viewers expected to perform.

\begin{figure*}[t]
\centering

\scalebox{0.75}{\input{key_hor2.tex}}\\\vspace{0.1cm}
\includegraphics[width=0.45\textwidth]{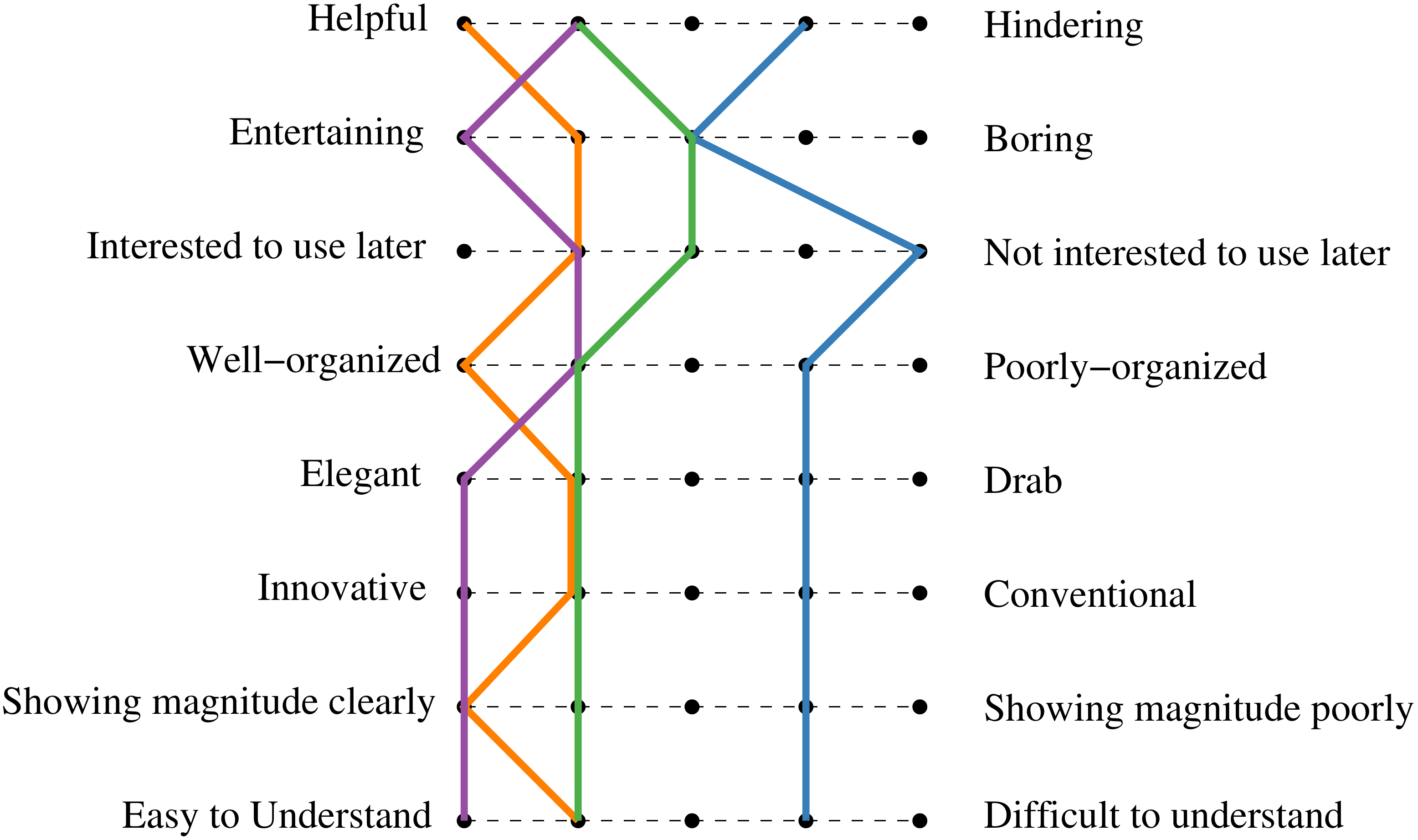}\hspace{.9cm}
\includegraphics[width=0.45\textwidth]{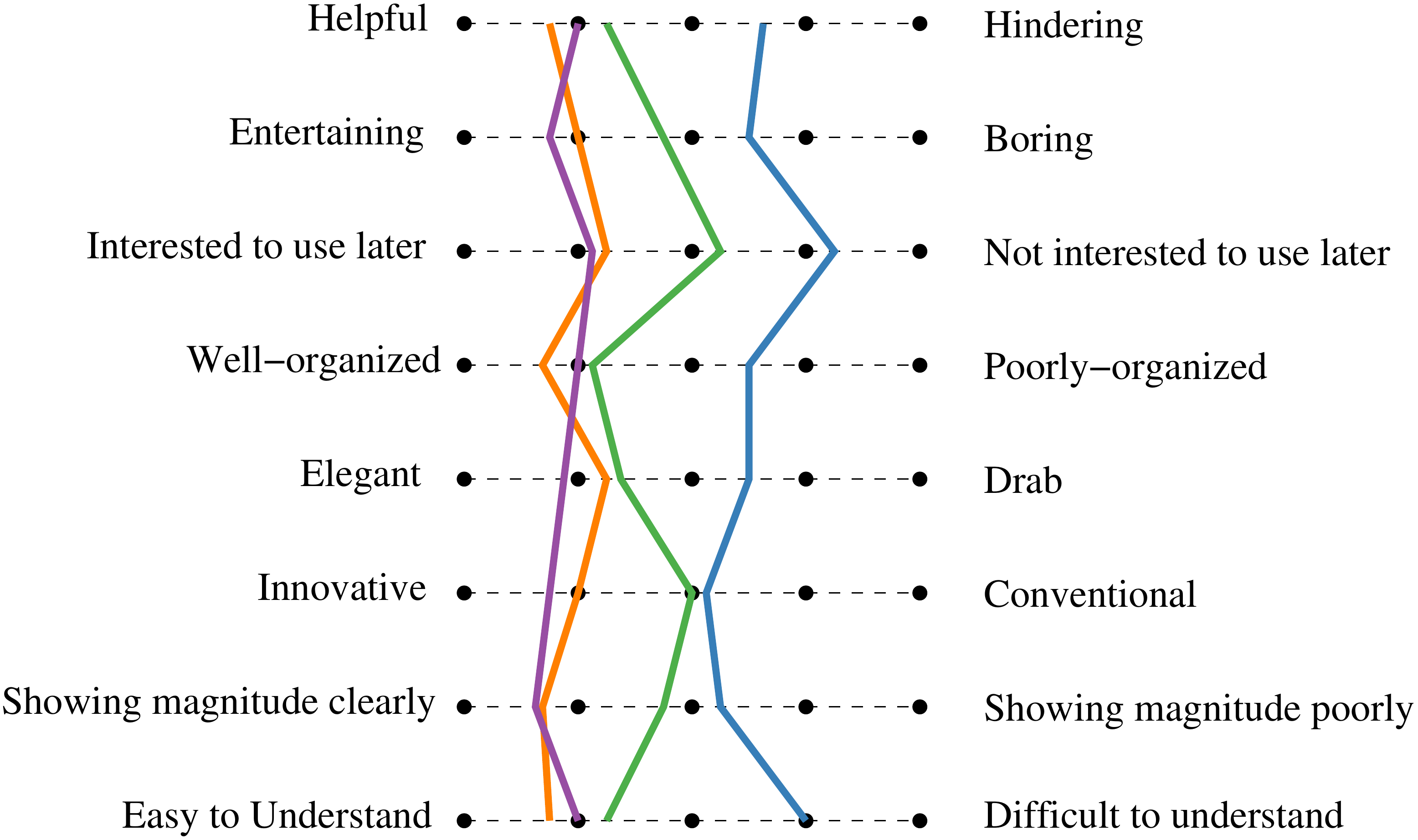}
\caption{
Subjective preference-based evaluation for four types of cartograms by mode (left) and mean (right): contiguous and Dorling cartograms clearly outperform the others~\cite{NusratAK15}.}
\label{fig:attitude}
\end{figure*}

\subsection{Subjective Preferences}

Sun and Li~\cite{Hui} analyzed the effectiveness of different types of cartograms by collecting subjective preferences. Two types of experimental tests were conducted: (1) a comparison between cartograms and thematic maps (choropleth maps, proportional symbol maps and dot maps), and (2) a comparison between cartograms (non-contiguous cartogram, diffusion cartogram, rubber-sheet cartogram, Dorling cartogram, and pseudo-cartogram). 
The participants in this study were  asked to select one map that is more effective for the representation of the given dataset and to provide reasons for this choice. 
The results indicate that cartograms are more effective in the representation of qualitative data (nominal data, such as, who won -- republican or democrats?), but thematic maps are more effective in the representation of quantitative results (ordinal data, such as, which region grew more?).
Note that for both experiments 
the subjects gave their preferences, but were not asked to perform any specific tasks.

Nusrat et al.~\cite{NusratAK15} also used subjective preferences, after performing several tasks, to evaluate four types of cartograms (contiguous, non-contiguous, rectangular, and Dorling). 
To understand user preferences about different aspects of cartograms, the semantic differential technique of Dent~\cite{dent1975} was used. Specifically, a rating scale between pairs of words or phrases that are polar opposites was used. There were five marks between these phrases and the participants selected the mark on a linear scale that best represented their attitudes for a given map and a given aspect. 
Participants were asked to rate the different cartogram types according to different categories, such as {\em helpfulness} of the visualization, {\em readability}, and {\em appearance}. The rank in each scale was constructed by calculating the mode (most frequent response) and the mean; see Fig.~\ref{fig:attitude}. The results indicated a clear preference for contiguous and Dorling cartograms. Specifically, the participants found contiguous cartograms to be helpful, well-organized and showing relative magnitude clearly, and Dorling cartograms to be entertaining, elegant, innovative, showing magnitude clearly, and easy to understand. The answers to the question ``Will you use this visualization later?'' also favored contiguous and Dorling cartograms.


Koletsis et al.~\cite{koletsis2014identifying} reported on early findings to identify possible approaches for evaluating the usability of different types of maps (e.g., nautical maps, pedestrian maps, and statistical maps). The aspects of map usability considered included: 
think aloud protocols, questionnaires, focus groups, participant feedback/formal and informal interviews, completion of map reading tasks, use of real and simulated environments, and statistical analyses for interpretation of results.
While studying effective ways to display geo-referenced statistical data, Pickle~\cite{pickle2003usability} devised a set of recommendations for statistical maps: the map should be designed for a particular audience and purpose, a standard legend should be used, colors should be chosen for the visually impaired and consistent with color conventions, and
multiple maps might be needed to address different questions. 
Although  Pickle considered choropleth maps and other thematic maps, most of these recommendations (with some extensions and modification) apply to cartograms.  

In order to improve cartogram design, Tao~\cite{Manting} conducted an online survey to collect views and suggestions from map viewers. 
The majority of the participants found cartograms to be difficult to understand and agreed that supporting references and help are indispensable in designing cartograms. Nevertheless, the survey did confirm that
 cartograms are commonly regarded as members of the map ``family''.

\subsection{Effectiveness of Interactivity in Cartograms}

Online cartograms are often interactive and animated. 
Ware~\cite{Jen} evaluated the effectiveness of animation in cartograms with a user-study in which ``locate'' and ``compare'' tasks were considered. The results indicate that although the participants preferred animated cartograms, the response time for the tasks was best in static cartograms. 

Cartographic interactions have been shown to promote visual thinking. Visualization of maps, exploration of data, finding anomalies and trends help generate insights~\cite{SARAIYA, NORTH06} about geographical phenomena~\cite{Roth}. Although many geo-visualization tools now provide interactive features, there are arguments in support of the ``less-is-more'' approach. For example, Jones et al.~\cite{Jones} claim that Philbruck's simplicity principle~\cite{Phil} should apply to the design of cartographic interactions, as well as to cartographic representations. Some recent studies indicate that increased level of cartographic interaction does not add value to the cartographic representation; to the contrary, the complete freedom of interaction may make  problem solving more difficult~\cite{DOU, Jones}. Therefore, although interaction with maps may help with visual analytics and insights, it also adds some level of complexity. Determining the balance between the value added by interaction and animation, and the resulting increase in complexity, remains an important open problem.

\section{Cartogram Criticism}

Even though cartograms have been used for more than a century, they continue to be criticized from several different directions and for several different reasons.

\noindent
\textbf{Cartograms are Difficult to Interpret without Additional Information:}
Dorling~\cite{dorling96} writes ``A frequent criticism of cartograms is that even cartograms based upon the same variable for the same areas of a country can look very different''. In the context of diffusion cartograms  Hennig et al.~\cite{hennig2009re}, write ``These maps are open to potential criticism when it comes to their informative value. One such criticism is the variation of the depicted topic within the territorial borders is not taken into consideration''.
Fotheringham et al.~\cite{Fother} write that Dorling maps ''can be hard to interpret without additional information to help the user locate towns and cities.'' 
While difficulty of interpretation varies between different types of cartograms, some of these issues can be addressed with good design principles (e.g., including the original map along with the cartogram).

\noindent
\textbf{Area Perception and Challenges for Cartograms:}
The impact of parameters such as area, color, and texture on visualization and understanding has been studied in visualization and cartography. This is relevant to cartograms as different algorithms generate different types of shapes (circles, rectangles, irregular polygons). Bertin~\cite{BERTIN83} was one of the first to provide systematic guidelines to test visual encodings. He evaluated visual variables according to their effectiveness for encoding nominal, ordinal, and quantitative data. He found that spatial position best facilitates graphical perception across all data types, while color hue ranks highly for nominal (categorical) data, but poorly for quantitative data. 
Cleveland and McGill~\cite{cleveland1984graphical} extended Bertin's work 
by conducting tests from psychology. In their perceptual experiments, subjects were shown charts and asked to compare the quantitative values of two marks by estimating what percentage the smaller value was of the larger. Their human-subjects experiments show significant accuracy advantage for position judgments over both length and angle judgments, which in turn proved to be better than area judgments. These test results were used to refine variables of visual encoding.
Stevens~\cite{Steven_law} modeled the mapping between the physical intensity of a stimulus and its perceived intensity as a power law. His study showed that subjects perceive length with minimal bias, but underestimate differences in area. This finding is further supported by Cleveland et al.~\cite{cleveland1982judgments}. In a more recent study, Heer and Bostock~\cite{JM10} investigated the accuracy of area judgment between rectangles and circles. They found that they have similar judgment accuracy, and both are worse than length judgments. 
They also found that when rectangles are drawn with aspect ratios in \{2/3, 1, 3/2\}, squares (aspect ratio 1) provide the worst accuracy. This supports earlier results about graphical comparison of bars, squares, circles, and cubes~\cite{croxton1932graphic}. 
These results are also consistent with the findings of ``judgment of size'' by Teghtsoonian~\cite{teghtsoonian1965judgment}, who found that there is a stronger correlation between actual area and apparent area for irregular polygons than for circles. 

Dent~\cite{dent1975} surveyed work on magnitude estimation, highlighting the tendency of humans to estimate lengths correctly, but underestimate areas and volumes. Perceptual tests led Flannery~\cite{flannery} to use apparent scaling of circles (rather than absolute scaling) to compensate for the underestimation. However, there are also strong arguments for absolute scaling. Tufte advises data encoding be truthful~\cite{ref:Tufte:1983a}: ``The representation of numbers, as physically measured on the surface of the graphic itself, should be directly proportional to the numerical quantities represented''. Krygier~\cite{perceptual_scaling} suggests that ``good legend design could eliminate the perceptual problem''. 
It is clear that there are non-trivial issues associated with area perception, while there are good suggestions for possible strategies to address  them (e.g., good legend design).

 \noindent
\textbf{Shape Distortion Makes it Hard to Realize Geography:}
Woodruff~\cite{Andyblog}, a cartographer, criticizes the ubiquitous red-blue US elections diffusion cartogram for a number of reasons: 

\hspace{0.01\textwidth}\parbox{0.45\textwidth}{\textit{``Topology preservation at the expense of shape: even if I know what a county looks like on a normal map, I'm going to have a hard time identifying it here... The bottom line is that many -- perhaps even most -- cartograms are essentially used for shock value, for the ``that's a different perspective!'' response, which is exactly what they get. Too frequently they can't stand as maps on their own. I think the election cartogram is only of use when it's next to an undistorted map. The best maps and graphics are those that tell their story clearly and elegantly, not those that simply evoke an emotional response.''  }}


In a public forum for cartography and design, Duffman~\cite{cartotalk}, also a cartographer, writes:

\hspace{0.01\textwidth}\parbox{0.45\textwidth}{\textit{
``a cartogram cannot work if people cannot recognize the geography. It no longer surprises/shocks/intrigues if we can't figure out where anything is and how much larger/smaller a place is than we expect. There's definitely a balance that needs to be struck. Good cartograms are still uncommon, I think. But they're worthwhile when done well.''
}} 



As already indicated by Duffman, while ``good cartograms are still uncommon'', they can be ``worthwhile when done well.'' Once again, better algorithms and good design principles can address some of these concerns.

\noindent
\textbf{Remarks:}
Cartograms are controversial and generate strong responses. Cartogram critics are numerous and criticisms of cartograms are common. Research shows that there are challenges with area perception in cartograms~\cite{dent1975}. The distortion of shapes in many cartograms makes it difficult to recognize the underlying geography. However, good design choices, such as legends, labels, and basic interaction techniques (such as linking and brushing with the original map) can address many of the common critiques.

\begin{figure}
{
\centering
\includegraphics[width=0.23\textwidth]{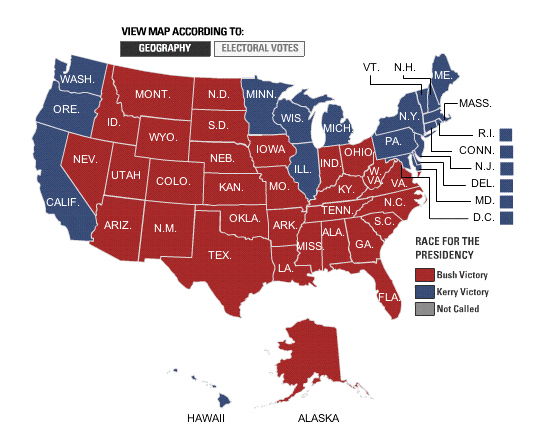} 
\includegraphics[width=0.24\textwidth]{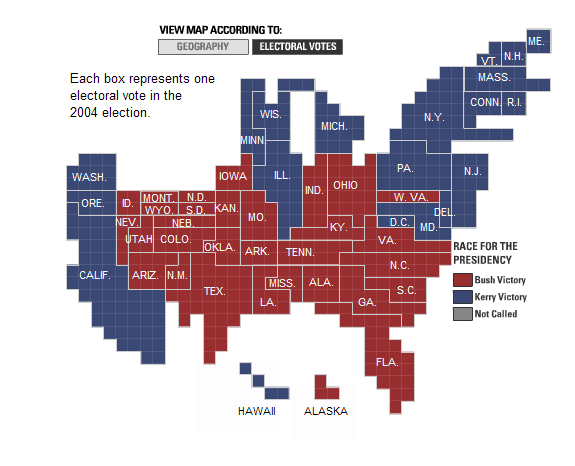}
\caption{Geographically accurate map and a population cartogram of the 2004 US election from the New York Times~\cite{NYT_04}.}
\label{fig:red-blue}
}
\end{figure}

\section{Applications}
\label{sec:application}
Unlike bar graphs (which represent size better), cartograms contain geographical information and adjacency relations. This makes it possible to see broader patterns and trends. In addition, cartograms make it possible to show more than one variable, e.g., population with size, winning candidate with color; see Fig.~\ref{fig:red-blue}. These are non-trivial advantages that make it possible to provide an overview and ``big-picture'' summary of the underlying data. As a result, cartograms are frequently used in scientific publications and in the popular press, in news report, blogs and presentations.

\subsection{Social Applications}
\label{Social}

\begin{figure}[htbp]
\begin{center}
\includegraphics[width=0.50\textwidth]{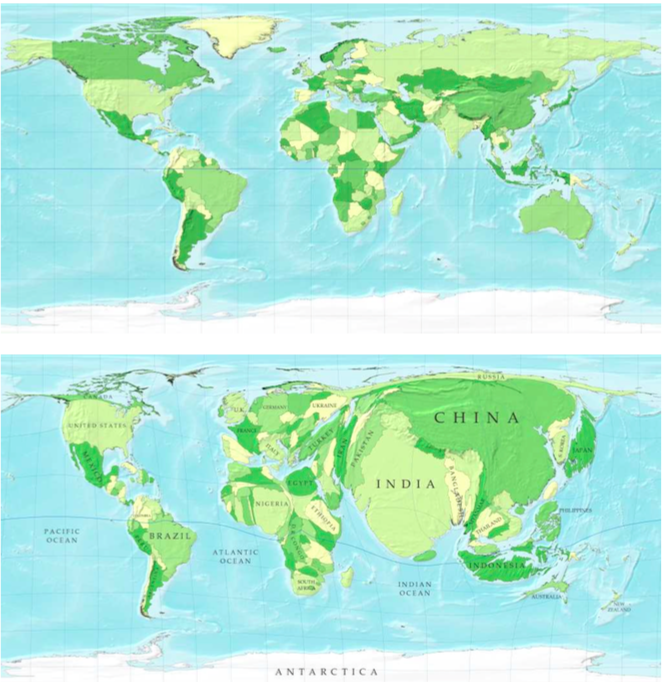}
\caption{Geographical world map (top), and population cartogram of the world (bottom)~\cite{map_week}.}
\label{fig:population}
\end{center}
\end{figure}

\begin{figure}[htbp]
\begin{center}
\includegraphics[width=0.48\textwidth]{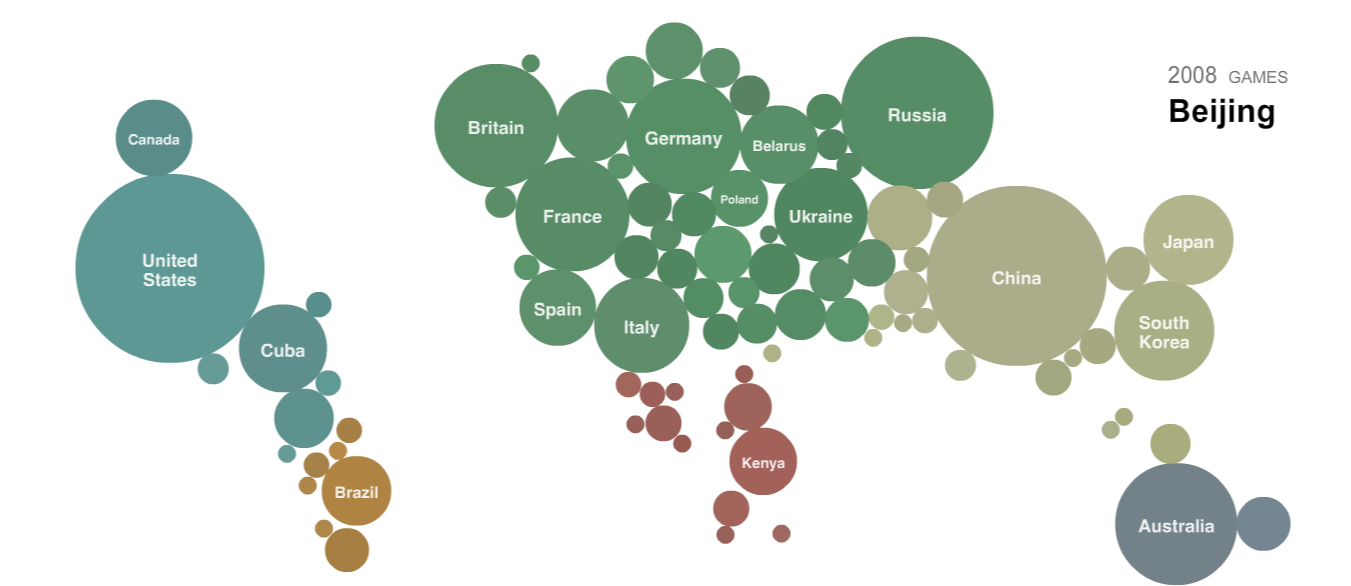}\\
(a)\\\vspace{.3cm}

\includegraphics[width=0.45\textwidth]{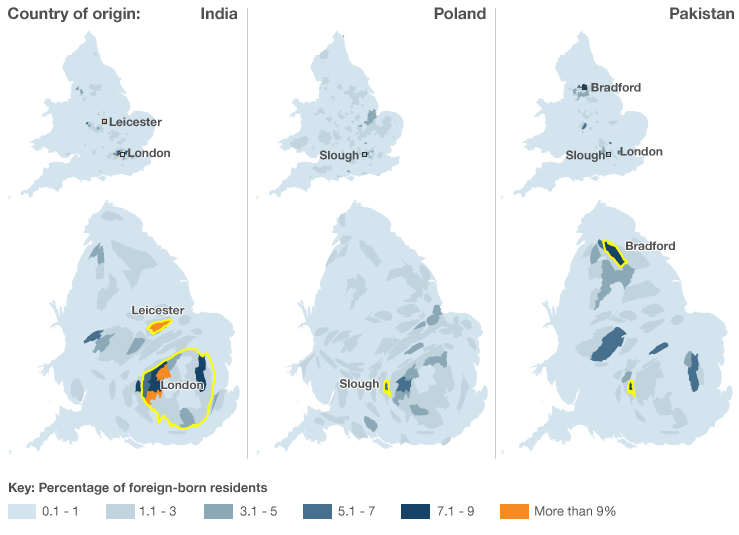}\\
(b)

\caption{(a) New York Times cartogram (Dorling) of the 2008 Olympic medals~\cite{NYT_O}; 
(b) 2012 BBC cartogram (diffusion) of migration patterns in the UK~\cite{BBCmigration}.
}
\label{fig:4Cart}
\end{center}
\end{figure}

A very common application of cartograms is to show population distribution; see Fig.~\ref{fig:population}. Dorling cartograms are used by the \textit{New York Times} to show the distribution of medals in the 2008 summer Olympic games~\cite{NYT_O}; see Fig.~\ref{fig:4Cart} (a). Similarly, Dorling cartograms are used by the \textit{Guardian}~\cite{Guar} in the UK to visualize social structure; see Fig.~\ref{fig:4Cart} (b).
Cartograms are also used to show the demographics of Twitter users~\cite{mislove2011understanding}, world citation and collaboration networks~\cite{pan2012world}, and wealth distribution in China~\cite{li2012cartograms}.

Popular TED talks use cartograms to illustrate how the news media can present a distorted view of the world~\cite{Alisa}, to highlight common misconceptions about the developing world~\cite{Hans2}, and to visualize the complex risk factors of diseases~\cite{Hans}. 
 Cartograms continue to be used in textbooks, for example, to teach middle-school and high-school students about global demographics and human development~\cite{Class1, Class2}.

\subsection{Political Applications}

\begin{figure}[htbp]
\begin{center}
\includegraphics[width=0.4\textwidth]{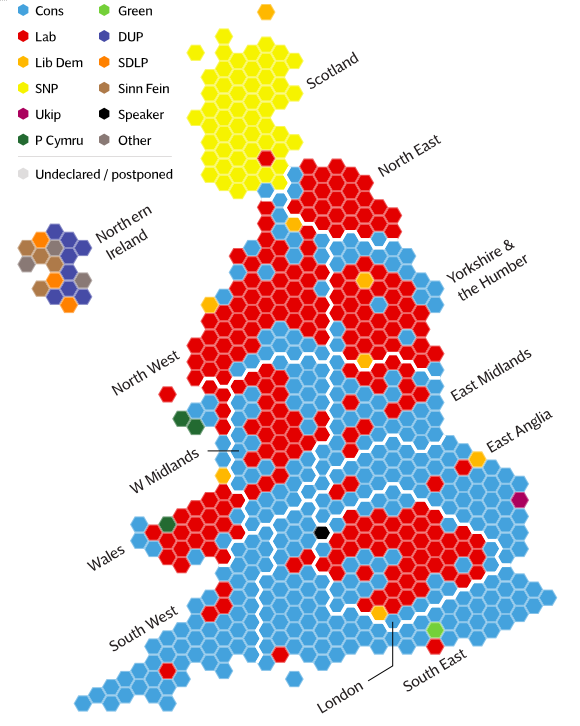}
\caption{Cartogram from The Telegraph showing the 2015 UK election results; each hexagon represents one seat in Parliament~\cite{UK15}.}
\label{fig:uk_e}
\end{center}
\end{figure}

Cartograms are frequently used to visualize election results. Despite the relative simplicity of choropleth maps, the media seems to prefer cartograms~\cite{Dailykos_vir}. For the last decade, the \textit{New York Times} has shown US election results with the now ubiquitous red-blue cartograms (e.g., in 2004~\cite{NYT_04} and in 2006~\cite{NYT06}). The \textit{Los Angeles Times}~\cite{LAT12} followed the trend using cartograms for the 2012 election results. 
Cartograms were also used to show the 2009 European Union election results of 2009 in the Dutch daily newspaper NRC~\cite{NRC}. The \textit{Telegraph} illustrated UK election results with a hexagonal cartogram~\cite{UK15}; see Fig.~\ref{fig:uk_e}. 

\subsection{Epidemiological Applications}

Disease mapping is used for tracking the spread of epidemics. Initially dot-maps and flow-maps were used, but since the mid-20th century more varied approaches became common~\cite{Manting}. 
The use of cartograms in disease mapping dates back to at least 1956 and has remained popular~\cite{bustamente1956geographical, levison1965area, hay2004global, bhatt2013global}. One reason for the appeal of cartograms in this setting is that the prevalence of a disease can seem more clustered than it actually is in a geographical map, while a cartogram can better show the distribution~\cite{kocmoud1997constructing}. 
The 1963 cartogram-like representation in ``National Atlas of Disease Mortality in the United Kingdom''~\cite{howe1963national} was motivated by the inaccurate visual representation of data density on geographical maps. 
The use of cartograms in the epidemiological domain grew with the rise of computer-generated cartograms. In 1986 Howe created a world cartogram of ``Global Geocancerology''~\cite{howe1986global} and in 1995, several disease cartograms were published in Dorling's ``New Social Atlas of Britain''~\cite{dorling1995new}.

\subsection{Other Applications}
 
\begin{figure}[htbp]
\begin{center}
\includegraphics[width=0.49\textwidth]{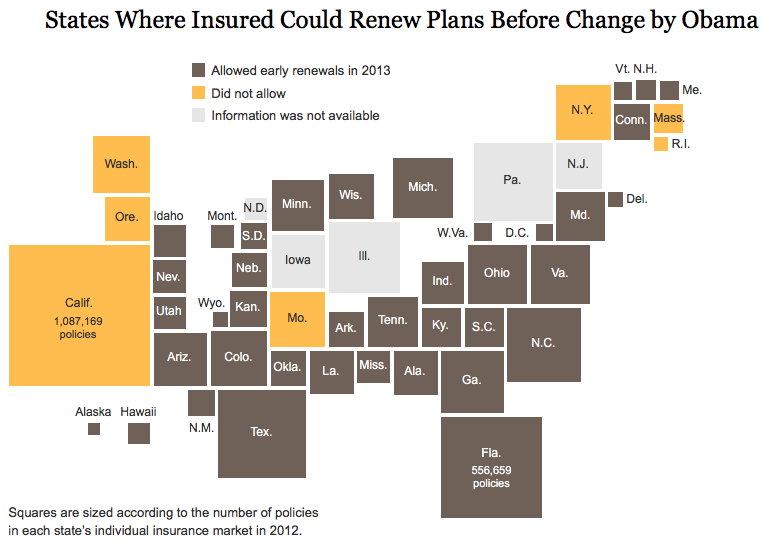}
\caption{A New York Times cartogram showing the number of individual health insurance plans per state~\cite{ins}.}
\label{fig:ins}
\end{center}
\end{figure}

Cartograms have been used in many different fields: to show distribution of grassland vegetation in Netherlands~\cite{de1956rough},  organic agriculture in the world~\cite{paull2013world}, global amphibian species studies~\cite{wake2008we}, product model visualization~\cite{tauscher2011area}, and even to show the changing face of global fisheries~\cite{watson2013changing}. 
Figure~\ref{fig:ins} shows a Demers cartogram from The \textit{New York Times} that resizes the US based on the number of {\em individual} health insurance plans in each state. Bruggmann et al.~\cite{bruggmann2013cartograms} used cartograms for visualization of non-geographic data, and user-generated content.


{

\begin{table}[t]
\caption{Tag categories for bibliography} 
\centering 
\begin{tabular}{ l   l   } 
\hline\hline 

Tag (category) & Description  \\ [0.5ex] 

\hline 

\textbf{\small type} & \parbox{0.3\textwidth}{\small \textbf{type of the paper}}\\[1.5ex]

\hspace{0.02\textwidth} {\small technique} & \parbox{0.3\textwidth}{\small novel technique for generating cartograms}\\[1.5ex]

\hspace{0.02\textwidth} {\small application} & \parbox{0.3\textwidth}{\small applying cartogram visualization to a specific application scenario}\\[1.5ex]

\hspace{0.02\textwidth} {\small evaluation} & \parbox{0.3\textwidth}{\small empirical, algorithmic, theoretical, or task-based evaluation} \\[1.5ex]

\textbf{\small cartogram\_type} & \parbox{0.3\textwidth}{\small \textbf{type of cartogram described in the paper}}\\[1.5ex]

\hspace{0.02\textwidth}  {\small generic} & \parbox{0.3\textwidth}{\small being applicable to all cartograms}\\[1.5ex]
		
\hspace{0.02\textwidth} {\small contiguous} & \parbox{0.3\textwidth}{\small being applicable to contiguous cartograms}\\[1.5ex]
		
\hspace{0.02\textwidth} {\small non-contiguous} & \parbox{0.3\textwidth}{\small being applicable to non-contiguous cartograms}\\[1.5ex]
		
\hspace{0.02\textwidth} {\small Dorling} & \parbox{0.3\textwidth}{\small being applicable to Dorling cartograms}\\[1.5ex]
		
\hspace{0.02\textwidth} {\small rectangular} & \parbox{0.3\textwidth}{\small being applicable to rectangular cartograms}\\[1.5ex]
		
\hspace{0.02\textwidth} {\small other} & \parbox{0.3\textwidth}{\small being applicable to some type of cartogram except the four major types}\\[1.5ex]

\textbf{\small evaluation} & \parbox{0.3\textwidth}{\small \textbf{kind of evaluation}}\\[1.5ex]

\hspace{0.02\textwidth} {\small algorithmic} & \parbox{0.3\textwidth}{\small testing the presented approach algorithmically or using metrics}\\[1.5ex]

\hspace{0.02\textwidth} {\small expert} & \parbox{0.3\textwidth}{\small assessing the approach through external domain or visualization experts}\\[1.5ex]

\hspace{0.02\textwidth} {\small none} & \parbox{0.3\textwidth}{\small no specific evaluation provided}\\[1.5ex]

\hspace{0.02\textwidth} {\small survey} & \parbox{0.3\textwidth}{\small specially broad survey of related work}\\[1.5ex]

\hspace{0.02\textwidth} {\small theoretical} & \parbox{0.3\textwidth}{\small theoretical considerations such as proof or runtime complexity}\\[1.5ex]

\hspace{0.02\textwidth} {\small user-study} & {\small conducting a study involving participants}\\[1.5ex]

\textbf{\small application} & \parbox{0.3\textwidth}{\small \textbf{area of application}}\\[1.5ex]

\hspace{0.02\textwidth} {\small social} & \parbox{0.3\textwidth}{\small social dynamics, population diversity and other data from social life}\\[1.5ex]

\hspace{0.02\textwidth} {\small epidemiology} & \parbox{0.3\textwidth}{\small prediction and distribution of diseases}\\[1.5ex]	

\hspace{0.02\textwidth} {\small political} & \parbox{0.3\textwidth}{\small political data such as election results}\\[1.5ex]			

\hspace{0.02\textwidth} {\small agriculture} & \parbox{0.3\textwidth}{\small agriculture-related data }\\[1.5ex]

\hspace{0.02\textwidth} {\small generic} & \parbox{0.3\textwidth}{\small no specific application suggested}\\[1.5ex]

\hline 
\end{tabular}
\label{table:tags} 
\vspace{-0.2cm}
\end{table}
}

\normalsize

\section{Bibliographic Analysis}
We used the SurVis web-based literature browser by Beck et al.~\cite{beck2016visual} for our bibliographic analysis. With the help of the system we analyzed cartogram papers published in journals and conferences in information visualization, cartography and computational geometry. We modified the system to incorporate different cartogram types and applications; the interactive cartogram literature browser is now available online~\cite{survis_carto}. The new tags and descriptions we used are summarized in Table~\ref{table:tags}.

\subsection{Historical Perspective}

Cartograms have a rich history dating back to the 19th century. At first, cartograms were manually created by geographers and cartographers~\cite{kasner1946distortion, hunter1968technique}. Soon statisticians, economists and journalists took interest, likely due to the possibility of creating appealing presentations of geo-referenced data. Early references to cartograms were mostly in the field of social studies~\cite{haro1968area, bustamente1956geographical}, agriculture~\cite{vinogradova1960agricultural, gerasimov1958geographical, de1956rough}, and epidemiology~\cite{levison1965area}. In the last couple of decades cartograms have been studied by computer scientists and cartographers. The growing number of publications, especially in theoretical computer science and visualization conferences, indicates an active research area.

Recent years saw the development of new cartogram-generation techniques, 
an increased interest in quantitative and qualitative evaluations, and their use in application domains. Figure~\ref{fig:tech-app} shows the growth trend in cartogram technique and application papers. 
For example, in EuroVis 2015, there were four publications, including a new  cartogram-generation technique~\cite{cano2015mosaic}, quantitative metrics for evaluating cartograms~\cite{AKV15}, a task taxonomy for cartograms~\cite{Task_C}, and an application of cartograms for visualizing the evolution of internet~\cite{johnson2015analyzing}.

It is worth mentioning that the SurVis system only keeps track of peer-reviewed publications. Thus many applications of cartograms in newspapers, magazines, and blogs are not considered here.

\subsection{Publications}

As suggested by Beck et al.~\cite{beck2016taxonomy}, we consider the number of citations (via Google Scholar) as a coarse quantifiable indicator of influence in academia. Table~\ref{table:tech_papers} shows the most cited cartogram-generation techniques. Note that, the most cited technique papers are from as early as 1934 to as late as 2007. The highest number of publications about cartograms occurred in 2013 with 16 papers: 6 of them were technique papers, 9 application papers, and 1 was an evaluation paper. Many application cartogram papers are also highly cited; see Table~\ref{table:app_papers}.

\begin{table}[h]
\caption{Most cited technique papers on cartograms} 
\centering 
\begin{tabular}{ l   l   l  l } 
\hline\hline 
Paper & Year & Cartogram type & Cit. \\ [0.5ex] 
\hline 

Gastner et al.~\cite{GN04} & 2004 & Contiguous & 411\\ [1ex] 
Dorling~\cite{dorling96} & 1996 & Dorling  & 161\\ [1ex] 

Dougenik et al.~\cite{DCN85} & 1985 & Contiguous & 122\\ [1ex] 

Raisz~\cite{Raisz34} & 1934 & Rectangular & 119\\[1ex] 

Keim et al.~\cite{KNP04} &  2004 & Contiguous & 99\\[1ex] 

van Kreveld et al.~\cite{ks07} & 2007 & Rectangular & 99\\[1ex] 

\hline 
\end{tabular}
\label{table:tech_papers} 
\vspace{-0.2cm}
\end{table}

\begin{table}[h]
\caption{Most cited papers on application of cartograms} 
\centering 
\begin{tabular}{ l   l  l l } 
\hline\hline 
Paper & Year & App type & Cit. \\ [0.5ex] 
\hline 

Bhatt et al.~\cite{bhatt2013global} & 2013  & epidemiology &1321\\ [1ex] 

Hay et al.~\cite{hay2004global} & 2004 &  epidemiology & 831\\[1ex] 

Wake et al.~\cite{wake2008we} & 2008 &  biology & 679\\ [1ex] 

Colizza et al.~\cite{colizza2006role} & 2006 & epidemiology & 579\\[1ex]

Mislove et al.~\cite{mislove2011understanding} & 2011 & epidemiology & 231\\ [1ex] 

\hline 
\end{tabular}
\label{table:app_papers} 
\vspace{-0.2cm}
\end{table}

\subsection{Topics and Trends}

\begin{figure*}[htbp]
\begin{center}
\includegraphics[width=0.7\textwidth]{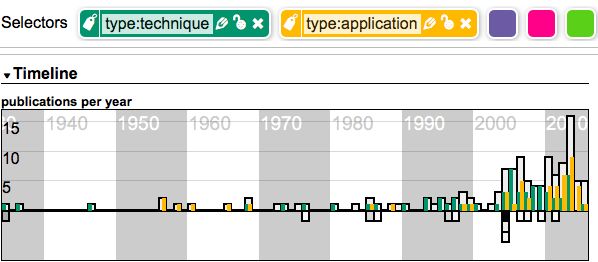}
\caption{Evolution of technique and application papers on cartograms. Downward bars
indicate the number of times a paper has been cited in that year.}
\label{fig:tech-app}
\end{center}
\end{figure*}

Based on our analysis, we observe the following trends:

\textbf{Growth in Application Papers:}
Figure~\ref{fig:tech-app} shows the number of technique and application papers over time. We can see that technique papers have been written at a steady pace over the last 30 years (with a spike in 2005), while application papers are noticeably growing over the last 15 years. We found several publications at highly cited journals, such as {\em Science} and {\em Nature}, that apply cartograms for studies in the diversity of amphibians on earth, the distribution of large social networks, and the spread of diseases.

\begin{figure*}[htbp]
\begin{center}
\includegraphics[width=0.7\textwidth]{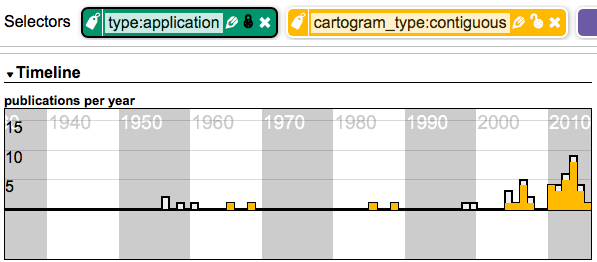}
\caption{Application papers with references to contiguous cartograms.}
\label{fig:app-cont}
\end{center}
\end{figure*} 

\begin{figure*}[htbp]
\begin{center}
\includegraphics[width=0.7\textwidth]{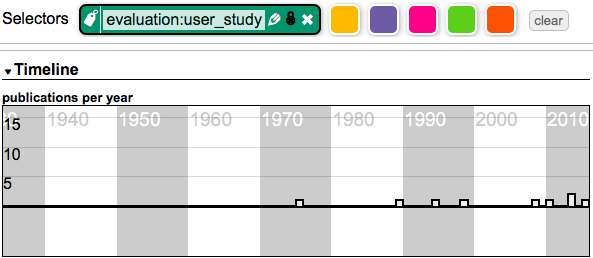}
\caption{Evaluation of cartograms.}
\label{fig:user-study}
\end{center}
\end{figure*}

\textbf{Contiguous Cartograms Dominate:}
From Figure~\ref{fig:app-cont}, we can see that contiguous cartograms are the most commonly used cartograms. Since the development of the diffusion-based method~\cite{GN04} in 2004, the adoption and application of this type of cartogram has been universal. This may be due to the availability of the software~\cite{carto_sw} to generate these cartograms, as well as their tendency to preserve shapes. Dorling cartograms are also easy to generate, and are used in abundance for web applications~\cite{Proto, Hans, Hans2, NYT_O, Guar}. However, scholarly articles still have a large bias towards diffusion cartograms.

\textbf{User-centric Evaluations:}
Cartograms are mostly evaluated based on pre-defined performance metrics. To understand the visual effectiveness and usability of cartograms and their applications, recently there has been some qualitative evaluation studies. From Figure~\ref{fig:user-study}, we can see that in the last decade several user-centric evaluations have been performed.  With new algorithms and cartogram visualizations on the rise, there is greater need to evaluate their communication effectiveness with task-based and qualitative evaluation. 

\subsection{State of the Field:}
The bibliographic analysis shows that cartogram visualization is an active and growing research area. There is a shift away from geography and cartography towards computational geometry and information visualization. There is also high demand for cartograms in many diverse applications. 

\section{Future Research Directions}

Recent evaluations suggest numerous design implications and promising research directions, especially due to the variety in cartogram types and in optimization goals. Here we summarize several possible research directions.

\subsection{Addressing Cartogram Limitations}






{
\begin{table*}[t]
\centering
\begin{tabular}{|c|c|c|c|c|c|}
\hline

\parbox{0.25\textwidth}{\centering \vspace{0.15cm}  \textbf{Type}

\vspace{0.15cm} } &

	\parbox{0.13\textwidth}{\centering \vspace{0.15cm} \textbf{Statistical accuracy}

\vspace{0.15cm} } &
	
	\parbox{0.13\textwidth}{\centering \vspace{0.15cm} \textbf{Topological accuracy}

\vspace{0.15cm} } &
	
	\parbox{0.13\textwidth}{\centering \vspace{0.15cm} \textbf{Geographical accuracy}}\\

\hline

\parbox{0.25\textwidth}{\centering \vspace{0.15cm} Contiguous cartograms \cite{GN04}

\vspace{0.15cm}
} &
	\parbox{0.13\textwidth}{\centering yes} &
	\parbox{0.13\textwidth}{\centering yes} &
	\parbox{0.13\textwidth}{\centering  \vspace{0.15cm} with small distortion \vspace{0.15cm}} \\

\hline

\parbox{0.25\textwidth}{\centering \vspace{0.15cm}  Dorling cartograms \cite{dorling96}

\vspace{0.15cm} 
} &
	\parbox{0.13\textwidth}{\centering yes} &
	\parbox{0.13\textwidth}{\centering no} &
	\parbox{0.13\textwidth}{\centering  no} \\

\hline

\parbox{0.25\textwidth}{\centering \vspace{0.15cm} Rectangular cartograms \cite{ks07}

\vspace{0.15cm} 
} &

	\parbox{0.13\textwidth}{\centering \vspace{0.15cm} depends on variant \vspace{0.15cm}} &

	\parbox{0.13\textwidth}{\centering depends on variant} &
	\parbox{0.13\textwidth}{\centering  no} \\

\hline

\parbox{0.25\textwidth}{\centering \vspace{0.15cm} Non-Contiguous cartograms \cite{Olson}

\vspace{0.15cm} 
} &
	\parbox{0.13\textwidth}{\centering yes} &
	\parbox{0.13\textwidth}{\centering no} &
	\parbox{0.13\textwidth}{\centering yes} \\

\hline

\end{tabular}
\caption{Performance of the four major types of cartograms on the three major design dimensions.}
\label{table:lim}

\end{table*}
}


 There are three major design dimensions along which cartograms may vary: statistical accuracy, geographical accuracy, and topological accuracy. We group cartogram-generation algorithms into four types (contiguous, Dorling, rectangular and non-contiguous) and summarize their performance on the three design dimensions; see Table~\ref{table:lim}. 
 
 Dorling cartograms can have zero cartographic error, but the shapes of the regions are not preserved. 
 When working with Dorling cartograms, the participants in a recent study~\cite{NusratAK15} had difficulties finding correct topological relations, even when the undistorted map was shown next to the cartogram. A similar problem has been identified for non-contiguous cartograms. In rectangular cartograms with zero cartographic error, there are usually large topological inaccuracies. However, simple interaction techniques may be used to mitigate such problems. 
 For example,  ``brushing'' that highlights the neighbors of a selected state can help identify the correct topological relations. 

Geographic inaccuracies occur in most cartogram types, with the notable exception of non-contiguous cartograms. A simple method that might alleviate geographic inaccuracies is to always show the cartogram alongside the undistorted geographical map. Linking and brushing between the map and the cartogram might provide sufficient hints to compensate for the shape and topology distortions.
 
 Finally, to deal with statistical inaccuracies and to mitigate the area estimation problem, the actual data value can be shown upon mouse-over events. Alternatively, a list of the data values can be shown next to the cartogram. 
 
Some of these techniques are well-known but they are rarely deployed in practice. Further, the effectiveness of these techniques in mitigating common cartogram problems has not been studied.
 



\subsection{Effectiveness of Other Cartogram Types}
Several studies have considered the effectiveness of the most popular types of cartograms. However, there are many new types of cartograms and variants of well-known cartograms. For example, the Dorling cartogram has several  variants, such as the square Demers cartogram~\cite{Bivar_book} and hexagonal Dorling cartogram~\cite{hennig2012rediscovering}, which have not been carefully examined. Similarly, new cartogram types such as mosaic cartograms~\cite{cano2015mosaic} and circular-arc cartograms~\cite{KKN13} have not been evaluated.

Most existing evaluations are quantitative evaluations of a small subset of cartograms, based on a couple of performance metrics such as cartographic error and shape distortion. While the performance metrics provide us the means for quantitative evaluation, we also need to evaluate the visual effectiveness of cartograms.
There are a handful of user-studies for cartograms, but few of them are broad and deep. More evaluation studies would benefit from including focus groups, interviews, think aloud protocols, questionnaires, and participant feedback. A good model for possible future work in this direction are cognitive and perception studies for geo-visualization, such as that by Fabrikant and Lobben~\cite{fabrikant2009introduction}. 

\subsection{Mapping Multi-variate Data}
\label{multivar}

\begin{figure}[htbp]
\begin{center}
\includegraphics[width=.49\textwidth]{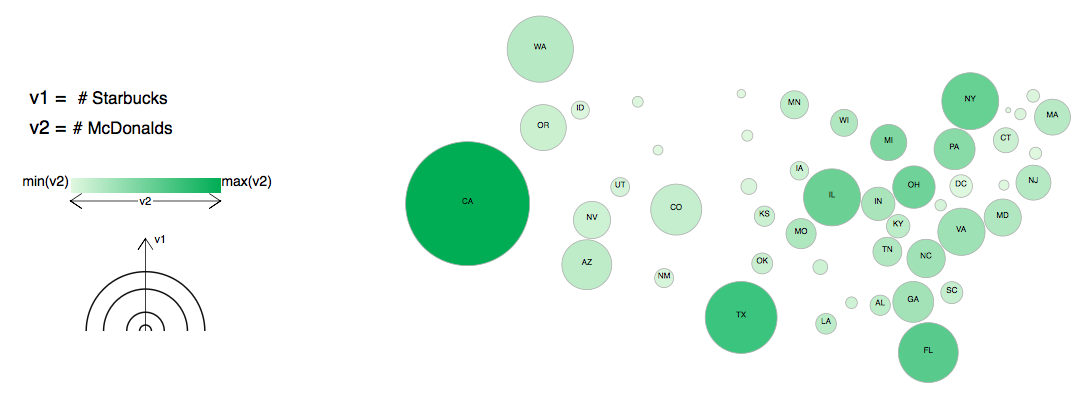}
\caption{A bivariate cartogram showing the number of McDonald's and Starbucks shops is the USA. The number of Starbucks shops is represented by circle size, and number of McDonald's shops is represented by color shading.}
\label{fig:choro}
\end{center}
\end{figure}

 Recently there has been some work on \textit{multivariate} map-based visualization~\cite{donnellan2004thematic,kaye2012mapping, FS04}. Identifying patterns and recognizing spatial relationships among multiple variables is an important feature of multivariate visualization. Cartograms are typically used to show one variable at a time. If more than one variable is used, the second variable is usually shown by color, as in choropleth maps~\cite{Bivar_book, bivar}; see Fig.~\ref{fig:choro}. The use of textures and glyphs is also common~\cite{carr1992hexagon, dorling1993computer, bivar2}. However showing too many variables on a map might make the visualization cluttered and hard-to-read~\cite{Bivar_book,kaye2012mapping}. The design, analysis and evaluation of multivariate cartograms is a promising research area.

\subsection{Memorability and Recall of Geographic Data}

We would like to create effective visualizations; visualizations that convey the most important and relevant aspects of the data or trend. In order to create effective data presentations, we first need to understand what makes a visualization memorable~\cite{borkin2013makes}. In~\cite{rittschof1994comparing}, the memorability (recall) of a cartogram was compared with that of the memorability of a choropleth map and other thematic maps. The participants were asked to redraw the maps/cartograms, and to estimate the data values in different regions from memory. 
In a follow-up study~\cite{rittschof1996learning}, the authors found that 
cartograms can be effective visualizations to recall geographical data, but recommended that cartograms be used only when the learners have a long-term familiarity with the depicted map.


Further evaluations on the memorability cartograms are needed. For example, the earlier studies~\cite{rittschof1998learning,rittschof1994comparing} used only one type of non-contiguous cartogram. There are no studies comparing the memorability of different types of cartograms. Another reason for further studies in this direction is the opportunity to evaluate the memorability of different types of cartograms using a spectrum of cartogram specific tasks~\cite{Task_C}.


\subsection{Uncertainty in Cartograms}

Uncertainty in geo-visualizations can occur for multiple reasons~\cite{maceachren2005visualizing}, such as:

\begin{itemize}
\item Lack of completeness (e.g., having the response rate less than 100\% in a survey)

\item Attribute inaccuracy (e.g., misunderstanding of the survey questions)

\item Spatial inaccuracy (e.g., address coding errors by the census enumerator)

\end{itemize}

Typical representation techniques for uncertain data on maps include contour crispness (crisp contours to show certain data, fuzzy contours to show uncertain data), focus (out-of-focus to show uncertain data), fill clarity (similar to focus), and color saturation. For example, MacEachren proposed that map elements with a high level of certainty should use pure hues, while those with less certain information should use a correspondingly less saturated color, thereby gray-ing out uncertain areas making their color hue ``uncertain''~\cite{maceachren1992visualizing}. 
Uncertain data representation has been studied in general for geo-visualization, but not specifically in the context of cartograms. 

\subsection{3D Cartograms}

 3D cartographic visualizations are studied both for technical and application purposes~\cite{KNP04, nollenburg2007geographic, WP-3D}. GIS tools allow us to create 3D cartograms~\cite{3d-carto1}. Little is known about the effectiveness of 3D cartograms, when compared to 2D cartograms, or when they are used to visualize multi-dimensional data.

\section{Conclusion}
We reviewed the state of the art in cartograms. Building on previous work in cartography and geography, cartograms have gained a great deal of attention in computational geometry and information visualization. Cartograms have been defined, generated, and classified in many different ways. 
By systemically collecting and categorizing the cartogram literature, we put together the different types of cartogram-generation algorithms, quantitative and qualitative metrics for evaluating cartograms, tasks and task taxonomies for cartograms, applications of cartograms, and future research directions.
With the growing interest in cartograms and cartogram applications, there is every indication that this will continue to be a productive research area. 

\section*{Acknowledgments}
We thank the authors of the SurVis system~\cite{beck2016visual} for allowing us to use it in preparing this survey.
This paper includes many images of cartograms. We thank the authors of the papers who granted us permission for the use of these images.

\newcommand{\etalchar}[1]{$^{#1}$}

\end{document}